# χ-sepnet: Deep neural network for magnetic susceptibility source separation

Minjun Kim[1], Sooyeon Ji[1,2], Jiye Kim[1], Kyeongseon Min[1], Hwihun Jeong[1], Jonghyo Youn[1], Taechang Kim[1], Jinhee Jang[3,4], Berkin Bilgic[5], Hyeong-Geol Shin[1,6,7,*], and Jongho Lee[1,*]

[1]Laboratory for Imaging Science and Technology, Department of Electrical and Computer Engineering, Seoul National University, Seoul, Republic of Korea

[2]Division of Computer Engineering, Hankuk University of Foreign Studies, Yongin, Republic of Korea

[3]Department of Radiology, Seoul St Mary's Hospital, College of Medicine, The Catholic University of Korea, Seoul, Republic of Korea

[4]Institute for Precision Health, University of California, Irvine, Irvine, CA, United States

[5]Massachusetts General Hospital and Harvard Medical School, Boston, MA, United States

[6]F.M. Kirby Research Center for Functional Brain Imaging, Kennedy Krieger Institute, Baltimore, MD, United States

[7]Russell H. Morgan Department of Radiology and Radiological Science, The Johns Hopkins University School of Medicine, Baltimore, MD, United States

**Correspondence**

Jongho Lee

Department of Electrical and Computer Engineering, Seoul National University, Seoul 08826, Republic of Korea

E-mail: jonghoyi@snu.ac.kr

Hyeong-Geol Shin

Department of Radiology, Johns Hopkins University School of Medicine, Baltimore, MD 21205, United States

E-mail: hshin40@jhu.edu

**Funding information**

This work was supported by the National Research Foundation of Korea (NRF-2022R1A4A1030579), NRF funded by the Korean government (MSIT) (RS-2024-00349509), Institute of Information & communications Technology Planning & Evaluation (IITP) under the artificial intelligence semiconductor support program to nurture the best talents (IITP-2023-RS-2023-00256081), INMC, Institute for the Study of New Media, IOER, Institute of Engineering Research, and AI-Bio Research at Seoul National University.

**Abstract**

Magnetic susceptibility source separation ($\chi$-separation), an advanced quantitative susceptibility mapping (QSM) method, enables the separate estimation of paramagnetic and diamagnetic susceptibility source distributions in the brain. Similar to QSM, it requires solving the ill-conditioned problem of dipole inversion, suffering from so-called streaking artifacts. Additionally, the method utilizes reversible transverse relaxation ($R_2' = R_2^* - R_2$) to complement frequency shift information for estimating susceptibility source concentrations, requiring time-consuming data acquisition for $R_2$ (e.g., multi-echo spin-echo) in addition to multi-echo GRE data for $R_2^*$. To address these challenges, we develop a new deep learning network, $\chi$-sepnet, and propose two deep learning-based susceptibility source separation pipelines, $\chi$-sepnet-$R_2'$ for inputs with multi-echo GRE and multi-echo spin-echo (or turbo spin-echo) and $\chi$-sepnet-$R_2^*$ for input with multi-echo GRE only. The neural network is trained using multiple head orientation data that provide streaking artifact-free labels, generating high-quality $\chi$-separation maps. The evaluation of the pipelines encompasses both qualitative and quantitative assessments in healthy subjects, and visual inspection of lesion characteristics in multiple sclerosis patients. The susceptibility source-separated maps of the proposed pipelines delineate detailed brain structures with substantially reduced artifacts compared to those from the conventional regularization-based reconstruction methods. In quantitative analysis, $\chi$-sepnet-$R_2'$ achieves the best outcomes followed by $\chi$-sepnet-$R_2^*$, outperforming the conventional methods. When the lesions of multiple sclerosis patients are assessed, both $\chi$-sepnet-$R_2'$ and $\chi$-sepnet-$R_2^*$ report identical lesion characteristics in most lesions ($\chi_{\text{para}}$: 99.6% and $\chi_{\text{dia}}$: 98.4% out of 250 lesions). The $\chi$-sepnet-$R_2^*$ pipeline, which only requires multi-echo GRE data, has demonstrated its potential to offer broad clinical and scientific applications, although further evaluations for various diseases and pathological conditions are necessary.

## 1. Introduction

In the brain, iron and myelin are the primary magnetic susceptibility sources, predominantly determining tissue susceptibility. These two substances have been shown to be involved in normal brain functions, and changes in their concentrations to be linked to the pathophysiology of various neurodegenerative disorders. For instance, iron deposition in substantia nigra is implicated in patients with Parkinson's disease (PD) (Dexter et al., 1987), and iron accumulation is also observed in the brains with Alzheimer's disease (AD) (Connor et al., 1992). Multiple sclerosis (MS) and Neuromyelitis Optic are known to introduce demyelination (Martino and Hartung, 1999; Wingerchuk et al., 2007).

Quantitative susceptibility mapping (QSM) is an MRI approach for quantifying bulk tissue magnetic susceptibility from magnetic field shift (or resonance frequency shift) (de Rochefort et al., 2008; Shmueli et al., 2009; Wang and Liu, 2015). Due to its sensitivity to paramagnetic iron and diamagnetic myelin, QSM has been utilized to monitor the brain pathology of neurodegenerative disorders associated with these susceptibility sources. In many brain substructures, however, iron and myelin co-exist within a voxel (Fukunaga et al., 2010), and influence QSM signal collectively. This makes it challenging to estimate their individual contributions to QSM, leading to a limited specificity for each substance.

To address this limitation, magnetic susceptibility source separation has been proposed, enabling separate estimation of the paramagnetic and diamagnetic susceptibility concentrations in brain tissue (Lee et al., 2017; Shin et al., 2021). The work by Shin et al. introduced $\chi$-separation (or chi-separation) technique to disentangle the contributions of para- and diamagnetic susceptibility sources ($\chi_{\text{para}}$ and $\chi_{\text{dia}}$) using both frequency shift and reversible transverse relaxation $R_2'$ ($= R_2^* - R_2$). This approach is based on the model where $R_2'$ is considered proportional to the absolute sum of paramagnetic and diamagnetic susceptibility concentrations (Lee et al., 2017; Shin et al., 2021). By successfully generating susceptibility source-separated maps, $\chi$-separation offers unique insights beyond traditional QSM, enabling the identification of MS lesions with iron and myelin alterations (Ji et al., 2024; Kim et al., 2023; Müller et al., 2024; Zhu et al., 2024) as well as visualization of cortical iron and myelin profiles (Lee et al., 2023). Similar susceptibility source separation approaches have been developed (Chen et al., 2021; Dimov et al., 2022; Emmerich et al., 2021; Kan et al., 2024; Li et al., 2023) and utilized in various applications (Jang et al., 2024; Nakashima et al., 2024). Chen et al. developed a multi-exponential model that only required multi-echo GRE data for susceptibility source separation (Chen et al., 2021). Dimov et al. proposed to approximate $R_2'$ from a linearly scaled $R_2^*$ to separate susceptibility sources without $R_2$ measurement (Dimov et al., 2022). Li et al. presented a method that suggested a voxel-specific parameter relating $R_2'$ to susceptibility (Li et al., 2023). Recently, methods have been proposed that utilize spatially adaptive relaxometric constants (Kan et al., 2024; Li et al., 2023).

Most of these studies either require multi-echo spin-echo (SE) for $R_2$ measurements, which is time-consuming, or rely on approximations that can introduce estimation errors (Ji et al., 2024). Additionally, similar to QSM, the susceptibility source-separated maps reconstructed from single-head orientation data may suffer from streaking artifact due to the ill-conditioned dipole inversion process (de Rochefort et al., 2008; Shmueli et al., 2009), even though regularization algorithms can reduce the artifacts to some extent (Li et al., 2015; Liu et al., 2011). Recently, studies have demonstrated that the application of deep learning algorithms effectively suppressed such artifacts and created high-quality maps (Bollmann et al., 2019; Yoon et al., 2018). Furthermore, deep learning has been successfully applied to MRI contrast synthesis when input contains information, such as structural details and image contrast, required to synthesize output (e.g., synthesizing STIR images from T1- and T2-weighted images) (Hagiwara et al., 2019; Moya-Sáez et al., 2021; Tanenbaum et al., 2023)

In this work, we present two χ-separation pipelines, χ-sepnet-$R_2'$ and χ-sepnet-$R_2^*$, that generate χ-separation maps using a deep neural network (χ-sepnet). The difference between the two pipelines is input: χ-sepnet-$R_2'$ demands both multi-echo GRE data and multi-echo SE data (or TSE), while χ-sepnet-$R_2^*$ requires only multi-echo GRE data. The missing $R_2'$ information in the χ-sepnet-$R_2^*$ pipeline is generated by another deep neural network, R2PRIMEnet, reconstructing an $R_2'$ map from an $R_2^*$ map. The proposed pipelines are validated using *in-vivo* data from both healthy subjects and MS patients to assess their feasibility for clinical utility.

## 2. Materials and methods

*2.1. Datasets for neural network training and evaluation*

*2.1.1. Data acquisition and processing for healthy subjects*

To construct χ-sepnet, a total of 72 scans was acquired from twelve healthy subjects (age = 25.5 ± 3.6 years; 8 males and 4 females) using a 3T MRI system (Siemens Tim Trio, Erlangen, Germany) equipped with a 32-channel phased-array head coil. All participants signed written consent approved by the institutional review board. Each subject underwent three-plane localization, manual $B_0$ shimming, magnetization-prepared rapid gradient echo (MPRAGE), 2D multi-echo SE, and then six 3D multi-echo GRE scans for different head orientations, starting with natural head orientation. For each GRE scan, the three-plane localization and manual $B_0$ shimming scan were re-conducted to ensure field homogeneity after head rotation. For all scans, the imaging slab was axial so that the z-axis was oriented along the $B_0$ field. Raw k-space data were collected from the scanner and reconstructed using in-house software. The total scan time was approximately 1 hour 10 minutes.

The MPRAGE sequence had the following parameters: repetition time (TR) = 2400 ms, echo time (TE) = 2.10 ms, field of view (FOV) = 256 × 256 × 224 mm$^3$, voxel size = 1 × 1 × 1 mm$^3$, flip angle = 8°, bandwidth = 210 Hz/pixel, GRAPPA factor = 2, and time of acquisition = 5.45 min. The 2D multi-echo SE data acquisition was performed using the following scan parameters: TR = 7800 ms, TEs = 15 ms, 30 ms, 45 ms, 60 ms, 75 ms, and 90 ms, FOV = 256 × 204 mm$^2$, number of slices = 76, slice gap = 0%, voxel size = 1 × 1 mm$^2$, slice thickness = 2 mm, flip angle = 90° with refocusing angle = 180°, bandwidth = 110 Hz/pixel, GRAPPA factor = 2, and time of acquisition = 13.25 min. The 3D multi-echo GRE data were acquired with the following parameters: TR = 38 ms, TEs = 7.70 ms, 12.73 ms, 17.76 ms, 22.79 ms, 27.82 ms, and 32.85 ms, FOV = 256 × 224 × 176 mm$^3$, voxel size = 1 × 1 × 1 mm$^3$, flip angle = 15°, bandwidth = 290 Hz/pixel, GRAPPA factor = 2, and time of acquisition = 6.42 min.

From each multi-echo GRE data, a local field map was calculated adhering to the QSM consensus guidelines (QSM Consensus Organization Committee et al., 2024). First, the multi-channel data were combined using sum-of-squares for the magnitude, and MCPC-3D for the phase (Robinson et al., 2011). A brain mask was then extracted from the combined magnitude image using FSL BET (Smith et al., 2004). Each phase image was unwrapped using ROMEO (Dymerska et al., 2021), followed by background field removal with V-SHARP, utilizing a spherical mean value size of 20 (Schweser et al., 2011; Wu et al., 2012b). Then, the background-removed multi-echo phase images were combined using the weighted echo averaging method to generate a local field map (Wu et al., 2012a). To create a QSM map from the multi-orientation GRE data, the local field maps from each orientation were registered to the first orientation data using FSL FLIRT (Smith et al., 2004), and the COSMOS QSM reconstruction was performed (Liu et al., 2009).

For $R_2^*$ mapping, auto-regression on linear operations (ARLO) (Pei et al., 2015) was utilized for multi-echo GRE data in each head orientation, enabling fast mono-exponential fitting. For $R_2$ mapping, a dictionary of spin-echo decay was simulated and matched for the multi-echo SE magnitude image with the stimulated echo correction (McPhee and Wilman, 2015). For the simulation, the StimFit toolbox (https://github.com/rmlebel/StimFit) was utilized. Subsequently, the $R_2^*$ map of each orientation and the $R_2$ map were registered to the first head orientation $R_2^*$ map using FSL FLIRT (Smith et al., 2004). The $R_2'$ maps were generated by subtracting the registered $R_2$ map from each orientational $R_2^*$ map. Any negative values were set to zero.

The registered multi-orientation $R_2'$ and local field maps were inputted to the multi-orientation $\chi$-separation algorithm ($\chi$-sep-COSMOS) (Shin et al., 2021), generating para- and diamagnetic susceptibility maps. The para- and diamagnetic susceptibility sources have positive and negative values, respectively (i.e., $\chi_{\text{para}} > 0$, $\chi_{\text{dia}} < 0$) but for simplicity, $-\chi_{\text{dia}}$ will be referred to as $\chi_{\text{dia}}$, hereafter. Any negative values in $\chi_{\text{para}}$ and $\chi_{\text{dia}}$ maps were set to zero to enforce physics. The relaxometric constant ($D_r$) was estimated by performing linear regression between the $R_2'$ and COSMOS QSM values in the five deep gray matter regions: caudate nucleus, putamen, globus pallidus, red nucleus, and substantia nigra. The resulting relaxometric contrast was 114 Hz/ppm with $R^2$ of 0.93, slightly deviating from the previously measured value of 137 Hz/ppm potentially due to different QSM algorithms (Shin et al., 2021).

*2.1.2. Data acquisition and processing for patients*

To evaluate the two deep learning pipelines in pathological conditions, MRI data from twelve MS patients (age: 32.3 ± 8.3 years; disease duration: 56.7 ± 35.6 months; five males and seven females; 11 patients treated with immunomodulatory drugs, one untreated), utilized in a previous study (Ji et al., 2024) were reprocessed.

The imaging protocol for the patient datasets included single-head orientation 3D multi-echo GRE, 2D dual-echo turbo spin-echo (DE-TSE), and 3D $T_2$-weighted FLAIR, acquired using a 3T MRI system (Magnetom Vida, Siemens Healthineers, Erlangen, Germany) equipped with a 64-channel phased-array head and neck coil. The 3D multi-echo GRE data were acquired with the following parameters: TR = 30 ms, TEs = 6.2 ms, 11.8 ms, 17.3 ms, and 22.9 ms, FOV = 187 × 230 × 144 mm³, voxel size = 0.7 × 0.7 × 1 mm³, flip angle = 18°, bandwidth = 270 Hz/pixel, GRAPPA factor = 2, and time of acquisition = 9.08 min. The 2D DE-TSE data acquisition was performed using the following scan parameters: TR = 11000 ms, TE = 10 ms, and 100 ms, FOV = 210 × 210 mm², number of slices = 120, slice gap = 0%, voxel size = 0.7 × 0.7 mm², slice thickness = 2 mm, flip angles = 90°, 165°, and 150°, bandwidth = 222 Hz/pixel, GRAPPA factor = 2, and time of acquisition = 6.05 min. The 3D $T_2$-weighted FLAIR sequence had the following parameters: TR = 7600 ms, TE = 431 ms, FOV = 230 × 230 × 180 mm$^3$, voxel size = 0.7 × 0.7 × 0.7 mm$^3$, flip angle = 90° with variable refocusing angles, bandwidth = 651 Hz/pixel, GRAPPA factor = 2 × 2, and time of acquisition = 2.56 min.

The multi-echo GRE data were processed to produce a local field map and $R_2^*$ map in accordance with the methods outlined in section 2.1.1. The DE-TSE data were processed for an $R_2$ map using dictionary matching with the StimFit toolbox (https://github.com/rmlebel/StimFit), assuming a normalized B1+ value of one throughout the brain. The $R_2$ maps of the patients were registered to the multi-echo GRE using FSL FLIRT (Smith et al., 2004). The $R_2'$ map was calculated by subtracting the $R_2$ map from $R_2^*$ map and subsequently setting negative values to zero.

For non-axial input data, the orientation was rotated such that the $B_0$ orientation is along the z-direction.

## 2.2. χ-sepnet-$R_2'$ and χ-sepnet-$R_2^*$ pipelines

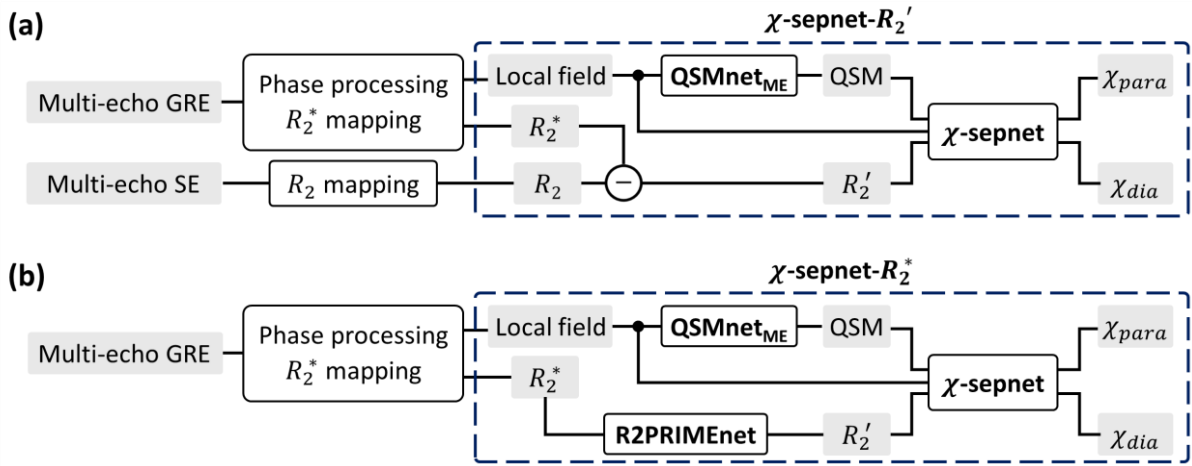

**Figure 1.** Overview of the proposed data processing pipelines for χ-separation with two different input options: χ-sepnet-$R_2'$ (a) and χ-sepnet-$R_2^*$ (b). (a) The χ-sepnet-$R_2'$ pipeline requires multi-echo GRE data for local field and $R_2^*$ maps, and multi-echo SE (or TSE) data for an $R_2$ map. (b) The χ-sepnet-$R_2^*$ pipeline is designed to operate solely on multi-echo GRE data. An $R_2'$ map is synthesized from $R_2^*$ using an additional neural network, R2PRIMEnet

The two processing pipelines, χ-sepnet-$R_2'$ and χ-sepnet-$R_2^*$, are shown in Figure 1. The two pipelines used the same data processing and deep learning-based χ-separation neural network (χ-sepnet), except for the procedures to estimate $R_2'$. The χ-sepnet-$R_2'$ pipeline consisted of the pre-processing steps of the multi-echo GRE and multi-echo SE data, producing local field, $R_2^*$ and $R_2$ maps. Then, the local field map was converted to a QSM map via a neural network, QSMnet$_{ME}$ (see Section 2.2.2). Finally, the QSM, local field, and $R_2'$ (= $R_2^*$ − $R_2$) maps were processed by χ-sepnet (see Section 2.2.1) to generate paramagnetic ($\chi_{para}$) and diamagnetic susceptibility maps ($\chi_{dia}$; positive-valued as mentioned in Section 2.1.1) as the output. For the $\chi$-sepnet-$R_2^*$ pipeline, only the GRE data were used as input, eliminating the need for the multi-echo SE data (Fig. 1b). After the pre-processing for local field and $R_2^*$ maps, QSM and $R_2'$ maps were generated by QSMnet$_{ME}$ and R2PRIMEnet, respectively. The latter was a newly designed network to extract $R_2'$ from $R_2^*$ (see Section 2.2.3). Finally, the QSM, local field, and $R_2'$ maps were processed via $\chi$-sepnet, creating $\chi_{\text{para}}$ and $\chi_{\text{dia}}$.

### 2.2.1. $\chi$-sepnet

$\chi$-sepnet was designed to reconstruct $\chi_{\text{para}}$ and $\chi_{\text{dia}}$ from the inputs of QSM, local field, and $R_2'$ maps. The three inputs instead of two inputs (i.e., local field and $R_2'$ maps) helped to apply the network for untrained resolution (see ablation study in Section 2.3). The three input maps were concatenated along the channel dimension. The output maps were designed to produce $\chi_{\text{para}}$ and $\chi_{\text{dia}}$ maps. For the network, a 3D U-net architecture, as utilized in QSMnet (Yoon et al., 2018), was employed with modifications to set the input and output channel numbers to 3 and 2, respectively (see Supplementary Section 1 for more details). Additionally, the output values were forced to be non-negative, setting any negative to zero.

For the training of $\chi$-sepnet, the $\chi$-sep-COSMOS maps from Section 2.1.1. were utilized as the label. The six healthy subjects (five for training, and one for validation) were utilized. From each of the six head orientations, an input and output pair was generated by registering the $\chi$-sep-COSMOS maps to the orientation of the input maps. Then, the pair was cropped into 3D patches with a size of $64 \times 64 \times 64$ for patch-based training. Each patch had 40 voxels of overlap with its adjacent neighbours. Data were augmented by rotating the input and output pair at six different angles per subject, creating 20,160 total patches for training after data augmentation. Unlike the augmentation in QSMnet, the rotation angle was applied to the plane perpendicular to $B_0$ (degree: -90° to +90°) to avoid complication from $B_0$ orientation dependent $R_2^*$ in white matter (Bender and Klose, 2010; Lee et al., 2011; Oh et al., 2013).

Before the training, the input and label data (e.g., local field in QSMnet$_{ME}$ and $\chi_{\text{para}}$ in $\chi$-sepnet) were normalized by calculating the mean and standard deviation values of all the training subjects within the brain mask. For each data, the normalization was performed by subtracting the mean value and then dividing by the standard deviation. The process was applied for all training, validation, and test datasets utilized for the neural networks.

The loss functions of the network were defined as follows: reconstruction loss ($\mathcal{L}_{recon}^{\chi sep}$), gradient loss ($\mathcal{L}_{gradient}^{\chi sep}$), and model loss ($\mathcal{L}_{model}^{\chi sep}$). To compute $\mathcal{L}_{recon}^{\chi sep}$, L1 loss was calculated between outputs and labels:

$$\mathcal{L}_{recon}^{\chi sep} = \left\| g(\chi, \varphi, R_2')_{para} - \chi_{para} \right\|_1 + \left\| g(\chi, \varphi, R_2')_{dia} - \chi_{dia} \right\|_1, \quad (1)$$

where $g(\cdot)_{para}$ and $g(\cdot)_{dia}$ denote the paramagnetic and diamagnetic susceptibility maps generated by $\chi$-sepnet, respectively. $\chi$, $\varphi$, and $R_2'$ refer to the QSM, local field, and $R_2'$ maps, respectively. $\chi_{para}$ and $\chi_{dia}$ are the labels that correspond to para- and diamagnetic susceptibility maps reconstructed by χ-sep-COSMOS, respectively. The gradient loss was designed to enhance the edge information, which was defined as:

$$\mathcal{L}^{\chi sep}_{gradient} = \left\| |\nabla g(\chi, \varphi, R'_2)_{para}|_x - |\nabla \chi_{para}|_x \right\|_1 + \left\| |\nabla g(\chi, \varphi, R'_2)_{dia}|_x - |\nabla \chi_{dia}|_x \right\|_1$$

$$+ \left\| |\nabla g(\chi, \varphi, R'_2)_{para}|_y - |\nabla \chi_{para}|_y \right\|_1 + \left\| |\nabla g(\chi, \varphi, R'_2)_{dia}|_y - |\nabla \chi_{dia}|_y \right\|_1$$

$$+ \left\| |\nabla g(\chi, \varphi, R'_2)_{para}|_z - |\nabla \chi_{para}|_z \right\|_1 + \left\| |\nabla g(\chi, \varphi, R'_2)_{dia}|_z - |\nabla \chi_{dia}|_z \right\|_1. \quad (2)$$

Lastly, the model loss enforced χ-sepnet to incorporate the physics information, represented by three losses ($\mathcal{L}^{\chi sep}_{QSM}$, $\mathcal{L}^{\chi sep}_{field}$, and $\mathcal{L}^{\chi sep}_{R'_2}$). These losses have similar ranges because the maps were normalized. The three losses were defined as:

$$\mathcal{L}^{\chi sep}_{QSM} = \left\| \left( g(\chi, \varphi, R'_2)_{para} - g(\chi, \varphi, R'_2)_{dia} \right) - \chi \right\|_1, \quad (3)$$

$$\mathcal{L}^{\chi sep}_{field} = \left\| \left( d * \left( g(\chi, \varphi, R'_2)_{para} - g(\chi, \varphi, R'_2)_{dia} \right) \right) - \varphi \right\|_1, \quad (4)$$

$$\mathcal{L}^{\chi sep}_{R'_2} = \left\| D_r \cdot \left( g(\chi, \varphi, R'_2)_{para} + g(\chi, \varphi, R'_2)_{dia} \right) - R'_2 \right\|_1, \quad (5)$$

$$\mathcal{L}^{\chi sep}_{model} = \mathcal{L}^{\chi sep}_{QSM} + \mathcal{L}^{\chi sep}_{field} + \mathcal{L}^{\chi sep}_{R'_2}, \quad (6)$$

where $d$ represents a dipole kernel, $*$ convolution operation, and $D_r$ a relaxometric constant. The final loss function was the weighted sum of the reconstruction, gradient, and model losses with the weights set to 1, 0.1, and 1, respectively.

For training, the RMSprop optimizer was utilized with a learning rate of 0.0003. The stepLR learning rate scheduler was employed with a step size of 1000 and a gamma value of 0.98. The network was trained with a batch size of 12. The training was terminated at 60 epochs because the performance reached a plateau. The final model was selected as the one with the lowest normalized root-mean-squared error (NRMSE) on the validation data. The training process was completed in approximately 47 hours.

### 2.2.2. QSMnet$_{ME}$

QSMnet$_{ME}$ is a neural network to generate a QSM map from a local field map. The network was designed using the same structure as QSMnet (Yoon et al., 2018) (see Supplementary Section 1). The primary difference between the QSMnet$_{ME}$ and QSMnet was the training data. QSMnet$_{ME}$ was trained using the multi-echo GRE data, which allows for phase offset compensation and provides a higher signal-to-noise ratio (Biondetti et al., 2020), whereas QSMnet was trained using single-echo GRE data.

The same subjects and approaches used for the χ-sepnet training were employed for the training of QSMnet_ME. Only the data augmentation was modified to match that of the QSMnet approach, by rotating the input and label maps at angles relative to $B_0$.

Three loss functions were utilized: L1 loss, gradient loss, and model loss, as in QSMnet+ (Jung et al., 2020). The loss weights for L1 loss, gradient loss, and model loss were set to 1, 0.1, and 1, respectively.

For training, the batch size was set to 12 and the learning rate was set to 0.001. The RMSprop optimizer and an exponential decay learning rate scheduler with a gamma value of 0.9999 were utilized. The training was conducted for 50 epochs, sufficient for the network's performance to converge, with the final model selected based on the minimum NRMSE on the validation data. The training time was approximately 31 hours.

When inferencing an input other than 1 mm isotropic resolution, which was the training data resolution for QSMnet_ME, a recently developed resolution generalization method was employed (Ji et al., 2023). This approach overcame the issue of resolution generalization in deep learning QSM (Jung et al., 2022), and enabled us to generate a QSM map up to the resolution of approximately 0.6 mm.

*2.2.3. R2PRIMEnet*

R2PRIMEnet was designed to estimate an $R_2'$ map from an $R_2^*$ map and was motivated by the previous contrast synthesis studies (Hagiwara et al., 2019; Moya-Sáez et al., 2021; Tanenbaum et al., 2023). The same network architecture as QSMnet_ME was deployed while setting the input and output to $R_2^*$ and $R_2'$, respectively (see Supplementary Section 1). The negative values of the output were set to zero.

The same subjects and approaches used for the χ-sepnet training were employed for the training of R2PRIMEnet.

The loss of R2PRIMEnet ($\mathcal{L}^{r2p}$) consists of L1 loss ($\mathcal{L}_1^{r2p}$) and gradient loss ($\mathcal{L}_{grad}^{r2p}$):

$$\mathcal{L}^{r2p} = \mathcal{L}_1^{r2p} + \mathcal{L}_{grad}^{r2p}, \quad (7)$$

$$\mathcal{L}_1^{r2p} = \|h(R_2^*) - R_2'\|_1, \quad (8)$$

$$\mathcal{L}_{grad}^{r2p} = \||\nabla h(R_2^*)|_x - |\nabla R_2'|_x\| + \||\nabla h(R_2^*)|_y - |\nabla R_2'|_y\| + \||\nabla h(R_2^*)|_z - |\nabla R_2'|_z\|. \quad (9)$$

where $h(\cdot)$ is R2PRIMEnet, $R_2^*$ is an $R_2^*$ map, and $R_2'$ is an $R_2'$ map. The total loss function was the weighted sum of $\mathcal{L}_1^{r2p}$ and $\mathcal{L}_{grad}^{r2p}$ with the weights set to 1 and 0.1 for $\mathcal{L}_1^{r2p}$ and $\mathcal{L}_{grad}^{r2p}$, respectively.

For training, the RMSprop optimizer was utilized with a learning rate set at 0.001. A stepLR learning rate scheduler was employed, featuring a step size of 1000 and a gamma value of 0.98. The network was trained with a batch size of 12, and the training was conducted for 50 epochs, which was sufficient for convergence. The total training time was approximately 30 hours. The final model was selected for the minimum NRMSE loss on the validation data.

*2.2.4. Implementation environments*

The data pre-processing methods for multi-echo GRE and multi-echo SE were conducted using MATLAB (MATLAB 2020a, MathWorks Inc., Natick, MA, USA). The networks were trained using PyTorch (Paszke et al., 2019) with two NVIDIA Quadro RTX-8000 GPUs.

*2.3. Experiments*

To evaluate the proposed pipelines in healthy subjects, the remaining six of the twelve healthy subjects were utilized. Two additional $R_2'$-based susceptibility source-separation methods ($\chi$-sep-MEDI and $\chi$-sep-iLSQR), which utilized MEDI regularization with a regularization parameter of 1 (Liu et al., 2011) and the iLSQR algorithm (Li et al., 2015), were processed (Shin et al., 2021) for comparison with the proposed pipelines. The evaluation of different algorithms was conducted by comparing the results to the $\chi$-sep-COSMOS maps as the reference. For quantitative evaluation, peak signal-to-noise (pSNR), normalized root-mean-squared error (NRMSE), high-frequency error norm (HFEN), and structural similarity (SSIM) were measured. All the metrics were calculated within the brain mask excluding cerebrospinal fluid (CSF) (Liu et al., 2018) and vessels (Kim et al., 2024).

A region of interest (ROI) analysis was conducted in a total of 24 brain regions, including gray matter and white matter regions. These ROIs were segmented from the $\chi$-separation atlas (see Supplementary Section 2) (Min et al., 2023).

An ablation study was designed to compare the performance of $\chi$-sepnet with that of a model excluding the QSM input (i.e., $\chi$-sepnet trained with field and $R_2'$ only) or the field input (i.e., $\chi$-sepnet trained with QSM and $R_2'$ only). All the networks had the same architecture as $\chi$-sepnet, except for the number of input channels (i.e., two input channels instead of three). The model loss term for the local field ($\mathcal{L}_{field}^{\chi sep}$) or for the QSM ($\mathcal{L}_{QSM}^{\chi sep}$) was not used for the network trained without local field or QSM, respectively.

An additional study was designed to assess the resolution generalization capability of $\chi$-sepnet, which was known to be an important issue in deep learning QSM (Jung et al., 2022). Since $\chi$-sepnet was trained using the 1 mm isotropic resolution data, inputs with different resolutions ($2 \times 2 \times 2$ mm$^3$, $1 \times 1 \times 3$ mm$^3$, and $0.7 \times 0.7 \times 0.7$ mm$^3$) were generated and tested. To generate the data with different resolutions, the k-space of the $1 \times 1 \times 1$ mm$^3$ input and label data used for $\chi$-sepnet test were cropped (for lower resolution) or zero-padded (for higher resolution). A 3D Tukey window with a filter size of 0.2 was then used to minimize artifacts in the resolution-modified maps. A resolution generalization method (Ji et al., 2023) was employed for QSMnet$_{ME}$. The performance of $\chi$-sepnet was compared with the two models utilized in the ablation study.

For the twelve MS patients, a qualitative assessment of lesions was conducted by a neuroradiologist (J. J.). Unlike the healthy subject data, the patient data were limited to a single-head orientation acquisition and, therefore, gold standard reference acquired from multi-orientation data did not exist. Hence, we only compared the consistency between the $\chi$-sepnet-$R_2'$ and $\chi$-sepnet-$R_2^*$ pipelines. MS lesions were segmented using a region growth algorithm from FLAIR images available at SPM12 software (https://www.fil.ion.ucl.ac.uk/spm/software/spm12/) (Egger et al., 2017) and fine-tuned by the neuroradiologist. A total of 250 lesions, each exceeding a volume of 20 mm$^3$, were evaluated for visual characteristics. The para- and diamagnetic susceptibility maps of the lesions were categorized into five classifications: hypointense, subtly hypointense, isointense, subtly hyperintense, or hyperintense relative to surrounding normal-appearing white matter (NAWM) (Kim et al., 2023).

## 3. Results

### 3.1. QSMnet$_{ME}$ and R2PRIMEnet in healthy subjects

The QSM maps generated by QSMnet$_{ME}$, along with the differences relative to the COSMOS QSM maps, are displayed in Figure 2a. The difference maps excluded CSF and vessels. As demonstrated in the figure, QSMnet$_{ME}$ successfully reconstructed COSMOS-quality QSM maps. The quantitative metrics for QSMnet$_{ME}$ (pSNR: 33.3 $\pm$ 0.9 dB, NRMSE: 48.0 $\pm$ 4.3%, HFEN: 36.8 $\pm$ 2.8%, SSIM: 0.950 $\pm$ 0.011; Fig. 2b) are slightly better than the literature values of QSMnet and QSMnet+ except for pSNR (see Discussion).

The $R_2'$ maps reconstructed by R2PRIMEnet and the difference images excluding CSF and vessels are displayed in Figure 2c. R2PRIMEnet effectively reconstructed the $R_2'$ maps from $R_2^*$ maps, showing little difference from the reference. The quantitative metrics for R2PRIMEnet (pSNR: 42.7 $\pm$ 4.2 dB, NRMSE: 25.3 $\pm$ 1.9%, HFEN: 44.9 $\pm$ 5.2%, SSIM: 0.949 $\pm$ 0.006; Fig. 2d) are better than those of QSMnet$_{ME}$ in terms of pSNR and NRMSE, while exhibiting comparable in SSIM.

These results demonstrate that QSMnet$_{ME}$ and R2PRIMEnet provide highly consistent maps with respect to the reference maps.

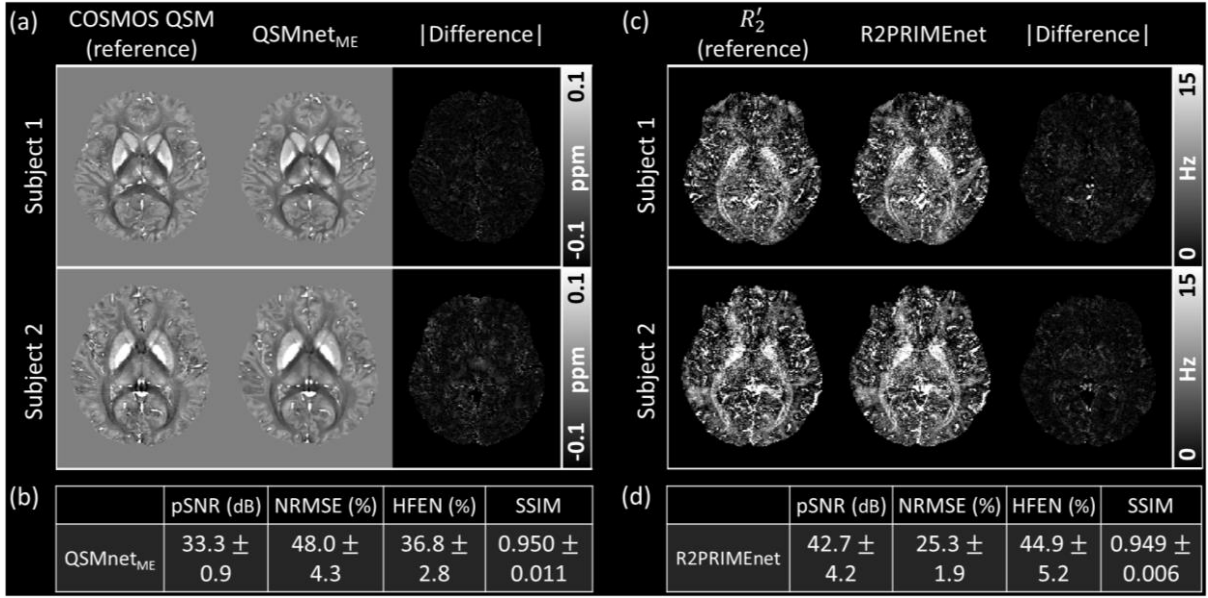

**Figure 2.** (a) QSM maps reconstructed by the COSMOS algorithm and QSMnet$_{ME}$ are shown alongside the absolute difference images. QSMnet$_{ME}$ successfully reconstructed COSMOS-quality QSM maps. (b) Quantitative metrics of QSMnet$_{ME}$ (pSNR, NRMSE, HFEN, and SSIM), calculated with respect to the COSMOS QSM maps. (c) $R_2'$ maps from MRI measurements (i.e., $R_2^* - R_2$) and R2PRIMEnet, along with the difference images. R2PRIMEnet effectively reconstructed the $R_2'$ maps, showing little differences from the reference. (d) Four metrics of R2PRIMEnet (pSNR, NRMSE, HFEN, and SSIM). pSNR and NRMSE of R2PRIMEnet are better than those from QSMnet$_{ME}$, while being comparable in SSIM. For pSNR and SSIM, higher values indicate superior performance, whereas for NRMSE and HFEN, lower values are indicative of better performance.

### 3.2. $\chi$-sepnet-$R_2'$ and $\chi$-sepnet-$R_2^*$ in healthy subjects

The $\chi_{\text{para}}$ and $\chi_{\text{dia}}$ maps reconstructed by five different $\chi$-separation reconstruction algorithms ($\chi$-sep-COSMOS, $\chi$-sepnet-$R_2'$, $\chi$-sepnet-$R_2^*$, $\chi$-sep-MEDI, and $\chi$-sep-iLSQR), along with the differences from the $\chi$-sep-COSMOS maps (as the reference), are displayed in Figure 3. The difference images are scaled by a factor of five and excluded CSF and vessels. Both $\chi_{\text{para}}$ and $\chi_{\text{dia}}$ maps from $\chi$-sepnet-$R_2'$ and $\chi$-sepnet-$R_2^*$ reveal less noise and higher correspondence to $\chi$-sep-COSMOS than those from $\chi$-sep-MEDI and $\chi$-sep-iLSQR, illustrating reduced discrepancies in the difference images. The quantitative metrics in Table 1 further consolidate these observations, with $\chi$-sepnet-$R_2'$ achieving the best outcomes, followed by $\chi$-sepnet-$R_2^*$.

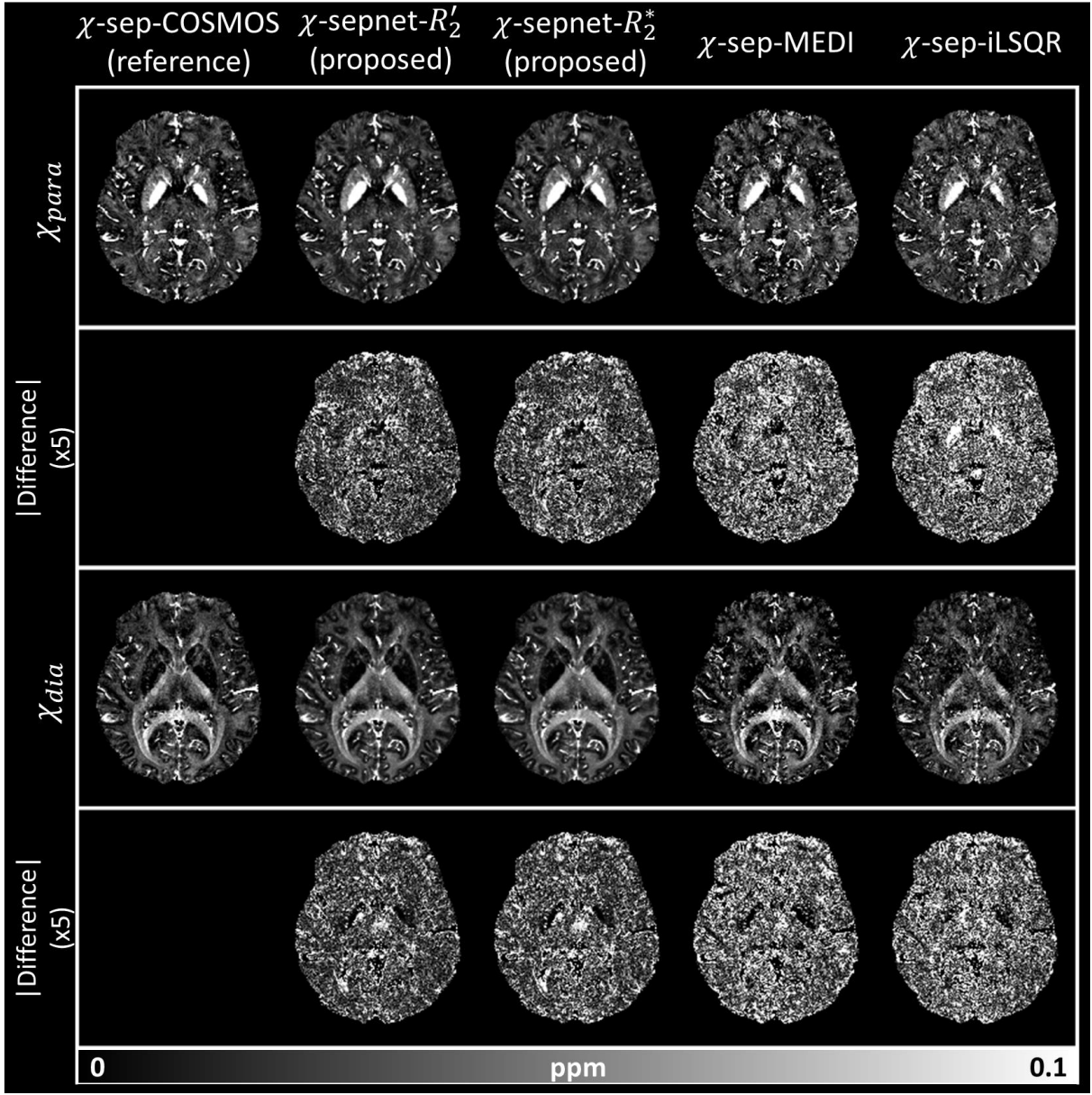

**Figure 3.** $\chi_{para}$ and $\chi_{dia}$ maps from the five $\chi$-separation methods (first row for $\chi_{para}$ and third row for $\chi_{dia}$) and their difference images to reference (second and fourth rows; scaled by a factor of five with CSF and vessels removed). The $\chi_{para}$ and $\chi_{dia}$ maps from the two proposed pipelines ($\chi$-sepnet-$R_2'$ and $\chi$-sepnet-$R_2^*$) display smaller differences compared to those from the other conventional algorithms ($\chi$-sep-MEDI and $\chi$-sep-iLSQR).

**Table 1.** pSNR, NRMSE, HFEN, and SSIM of the four $\chi$-separation algorithms, calculated with respect to $\chi$-sep-COSMOS.

|  |  | $\chi$-sepnet-$R_2'$ (proposed) | $\chi$-sepnet-$R_2^*$ (proposed) | $\chi$-sep-MEDI | $\chi$-sep-iLSQR |
|---|---|---|---|---|---|
| $\chi_{para}$ | pSNR (dB) | 39.1 $\pm$ 1.0† | 38.4 $\pm$ 0.8‡ | 35.2 $\pm$ 1.2 | 35.8 $\pm$ 1.0 |
|  | NRMSE (%) | 38.1 $\pm$ 2.8† | 41.6 $\pm$ 2.4‡ | 60.0 $\pm$ 6.8 | 55.8 $\pm$ 5.5 |
|  | HFEN (%) | 47.2 $\pm$ 4.2† | 55.3 $\pm$ 5.1‡ | 74.2 $\pm$ 8.0 | 63.9 $\pm$ 6.9 |
|  | SSIM | 0.940 $\pm$ 0.007† | 0.927 $\pm$ 0.007‡ | 0.868 $\pm$ 0.015 | 0.877 $\pm$ 0.014 |
| $\chi_{dia}$ | pSNR (dB) | 39.5 $\pm$ 0.9† | 38.6 $\pm$ 0.8‡ | 35.2 $\pm$ 1.2 | 35.9 $\pm$ 1.0 |
|  | NRMSE (%) | 35.7 $\pm$ 2.2† | 39.4 $\pm$ 1.8‡ | 60.0 $\pm$ 6.8 | 54.2 $\pm$ 5.0 |
|  | HFEN (%) | 48.4 $\pm$ 2.7† | 57.7 $\pm$ 2.1‡ | 78.0 $\pm$ 4.7 | 67.1 $\pm$ 4.4 |
|  | SSIM | 0.937 $\pm$ 0.008† | 0.922 $\pm$ 0.008‡ | 0.854 $\pm$ 0.017 | 0.864 $\pm$ 0.016 |

† Best metric among the four $\chi$-separation algorithms
‡ Second best metric among the four $\chi$-separation algorithms

When the $\chi_{para}$ and $\chi_{dia}$ maps are zoomed-in for details in basal ganglia region, including caudate nucleus (CN), putamen (Put), globus pallidus (GP), pulvinar (Pul), dorsomedial nucleus (DN), posterior limb of internal capsule (PLIC) and anterior limb of internal capsule (ALIC), they reveal hyperintense $\chi_{para}$ in iron-rich regions (CN, Put, GP, and Pul), hyperintense $\chi_{dia}$ in white matter regions (PLIC and ALIC), and hypointense $\chi_{dia}$ in regions known to have little myelin (CN, Put, GP, Pul, and DN), agreeing with previous histological observations (Naidich et al., 2012; Schaltenbrand, 1977) (Fig. 4a). Figure 4b shows a zoomed-in axial slice of the $\chi$-separation maps in the central sulcus region. The hand knob region clearly demonstrates a higher $\chi_{para}$ intensity in the motor cortex (MC) than that in the sensory cortex (SC), which is consistent with the well-known iron distribution observed in histology (Stuber et al., 2014). The cortical boundary appears blurred in the results from $\chi$-sep-MEDI and $\chi$-sep-iLSQR, potentially due to the streaking artifacts. Overall, the $\chi$-sepnet-$R_2'$ and $\chi$-sepnet-$R_2^*$ maps exhibit reduced noise compared to those from the other methods. In particular, the maps from $\chi$-sep-COSMOS are also noisier than those from $\chi$-sepnet-$R_2'$ and $\chi$-sepnet-$R_2^*$ partially because of registration issues (see Discussion). Vessels show up in both $\chi_{para}$ and $\chi_{dia}$, which is an artifact in $\chi$-separation (Lee et al., 2024).

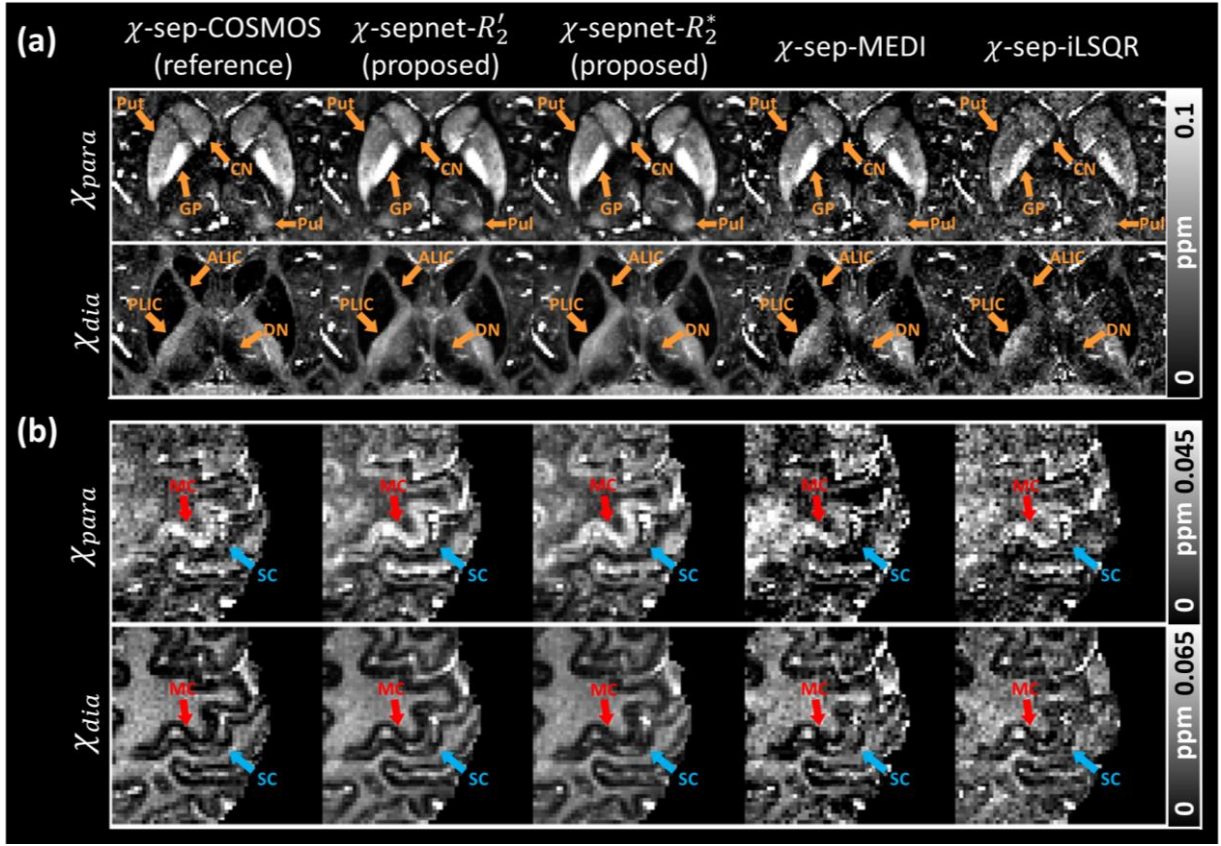

**Figure 4.** Zoom-in axial slice of the $\chi$-separation methods, displaying basal ganglia (a) and central sulcus (b). (a) Deep gray matter regions (Put: putamen, CN: caudate nucleus, GP: globus pallidus, Pul: pulvinar, and DN: dorsomedial nucleus) and white matter regions (ALIC: anterior limb of internal capsule, and PLIC: posterior limb of internal capsule) are clearly delineated in the $\chi_{para}$ and $\chi_{dia}$ maps. (b) In the hand-knob region, maps from $\chi$-sepnet-$R_2'$ and $\chi$-sepnet-$R_2^*$ manifest a higher concentration of positive susceptibility sources (most likely iron) in the motor cortex (MC) than in the sensory cortex (SC). Vessels show up in both $\chi_{para}$ and $\chi_{dia}$, which is an artifact in $\chi$-separation.

The results of the ROI analysis in the healthy subjects are presented in Figure 5. The $\chi$-sepnet-$R_2'$ and $\chi$-sepnet-$R_2^*$ pipelines exhibit strong quantitative correspondence to the reference ($\chi$-sep-COSMOS), reporting slopes of 1.043 and 1.051 for $\chi_{para}$, respectively, and 0.939 and 0.923 for $\chi_{dia}$, respectively. The results from $\chi$-sep-MEDI also exhibit good consistency with the reference, showing slopes of 0.938 for $\chi_{para}$ and 0.934 for $\chi_{dia}$, whereas $\chi$-sep-iLSQR reveals the lowest correspondence (slopes of 0.835 and 0.813 for $\chi_{para}$ and $\chi_{dia}$, respectively). Similarly, the $R^2$ values demonstrate the best correlations of $\chi$-sepnet-$R_2'$ with respect to the reference, with $R^2$ of 0.994 and 0.974 for $\chi_{para}$ and $\chi_{dia}$, respectively. $\chi$-sepnet-$R_2^*$ also shows good correlations with the reference with $R^2$ of 0.993 for $\chi_{para}$ and 0.966 for $\chi_{dia}$. In contrast, $\chi$-sep-MEDI and $\chi$-sep-iLSQR reveal inferior performances, reporting $R^2$ of 0.989, and 0.984 for $\chi_{para}$, respectively, and 0.955, and 0.914 for $\chi_{dia}$, respectively. All measurements of the 24 ROIs are listed in Supplementary Section 2.

All the results so far suggest that $\chi$-sepnet-$R_2'$ provides the best performance among the four methods and is closely followed by $\chi$-sepnet-$R_2^*$. On the other hand, the other two methods reveal qualitative and quantitative performance degradation.

The results of the ablation study and the resolution generalization study show that the proposed model with all three inputs performs the best while the model, trained with QSM and $R_2'$ maps, matches the performance (see Supplementary Sections 3 and 4). On the other hand, the model, trained with local field and $R_2'$ maps, demonstrates poor performance in the resolution generalization study for anisotropic resolution ($1 \times 1 \times 3$ mm$^3$) (see Supplementary Section 4).

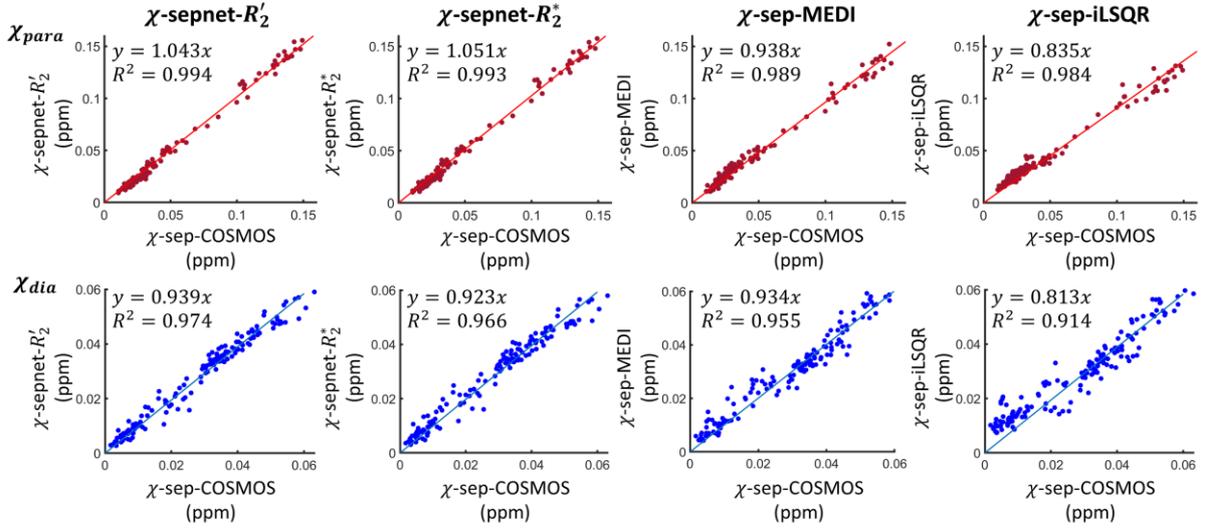

**Figure 5.** ROI analysis results. Each point in the plot represents the mean value of one of the 26 ROIs in one participant. The linear regression slope and $R^2$ value are displayed in each panel, and reveal that $\chi$-sepnet-$R_2'$ has the best performance, closely followed by $\chi$-sepnet-$R_2^*$. $\chi$-sep-iLSQR shows the poorest performance.

### 3.3. $\chi$-sepnet-$R_2'$ and $\chi$-sepnet-$R_2^*$ in MS patients

When $\chi_{\text{para}}$ and $\chi_{\text{dia}}$ are reconstructed by $\chi$-sepnet-$R_2'$ and $\chi$-sepnet-$R_2^*$, and their lesion appearance are compared in the MS patients, majority of lesions exhibit similar contrasts. In terms of lesion categorization based on $\chi_{\text{para}}$ and $\chi_{\text{dia}}$ contrasts, $\chi_{\text{para}}$ maps show consistent results in 249 out of the 250, while $\chi_{\text{dia}}$ maps do in 246 lesions (Tables 2 and 3). Most lesions appear isointense or hyperintense compared to surrounding NAWM in $\chi_{\text{para}}$ (isointense: 137, hyperintense: 96; Table 2), while $\chi_{\text{dia}}$ typically shows hypointense or subtle hypointense lesions (hypointense: 159, subtle hypointense: 44; Table 3). No lesion exhibits hyperintense or even subtle hyperintense in $\chi_{\text{dia}}$. These findings are consistent with the fact that demyelination is one of the primary pathophysiology of MS (Kim et al., 2023).

In Figure 6, representative MS lesions from each category are visualized in $T_2$-weighted FLAIR image, and $\chi_{\text{para}}$ and $\chi_{\text{dia}}$ maps, with yellow arrows indicating lesion locations. The blue boxes reveal the results with good consistency between $\chi$-sepnet-$R_2'$ and $\chi$-sepnet-$R_2^*$. The red boxes specify the results with the discrepancy, where the lesions appear subtly hypointense in the $\chi_{para}$ map of $\chi$-sepnet-$R_2'$, but hypointense in $\chi$-sepnet-$R_2^*$ (Fig. 6f), and subtly hypointense in the $\chi_{dia}$ maps of $\chi$-sepnet-$R_2'$, but hypointense (Fig. 6j) and isointense (Fig. 6k) in $\chi$-sepnet-$R_2^*$.

Note that only five out of 500 lesions (both $\chi_{para}$ and $\chi_{dia}$ lesions combined) are categorized differently between the two methods. Furthermore, the differences are subtle as demonstrated in Figs. 6f, j, and k. These outcomes suggest that $\chi$-sepnet-$R_2^*$ can successfully generate $\chi$-separation results without a $R_2$ map when evaluating $\chi$-separation characteristics in MS lesions.

**Table 2.** Total of 250 MS lesions are assessed for visual characteristics in the $\chi_{para}$ maps from the $\chi$-sepnet-$R_2'$ and $\chi$-sepnet-$R_2^*$ pipelines.

| $\chi_{para}$ | | $\chi$-sepnet-$R_2^*$ | | | | | |
|---|---|---|---|---|---|---|---|
| | | Hyper | Subtle hyper | Iso | Subtle hypo | Hypo | Total |
| $\chi$-sepnet-$R_2'$ | Hyper | 96 | 0 | 0 | 0 | 0 | 96 |
| | Subtle hyper | 0 | 1 | 0 | 0 | 0 | 1 |
| | Iso | 0 | 0 | 137 | 0 | 0 | 137 |
| | Subtle hypo | 0 | 0 | 0 | 5 | 1 | 6 |
| | Hypo | 0 | 0 | 0 | 0 | 10 | 10 |
| | Total | 96 | 1 | 137 | 5 | 11 | 250 |

**Table 3.** Total of 250 MS lesions are evaluated for their visual characteristics using the $\chi_{dia}$ maps generated by the $\chi$-sepnet-$R_2'$ and $\chi$-sepnet-$R_2^*$ pipelines.

| $\chi_{dia}$ | | $\chi$-sepnet-$R_2^*$ | | | | | |
|---|---|---|---|---|---|---|---|
| | | Hyper | Subtle hyper | Iso | Subtle hypo | Hypo | Total |
| $\chi$-sepnet-$R_2'$ | Hyper | 0 | 0 | 0 | 0 | 0 | 0 |
| | Subtle hyper | 0 | 0 | 0 | 0 | 0 | 0 |
| | Iso | 0 | 0 | 43 | 0 | 0 | 43 |
| | Subtle hypo | 0 | 0 | 1 | 44 | 3 | 48 |
| | Hypo | 0 | 0 | 0 | 0 | 159 | 159 |
| | Total | 0 | 0 | 44 | 44 | 162 | 250 |

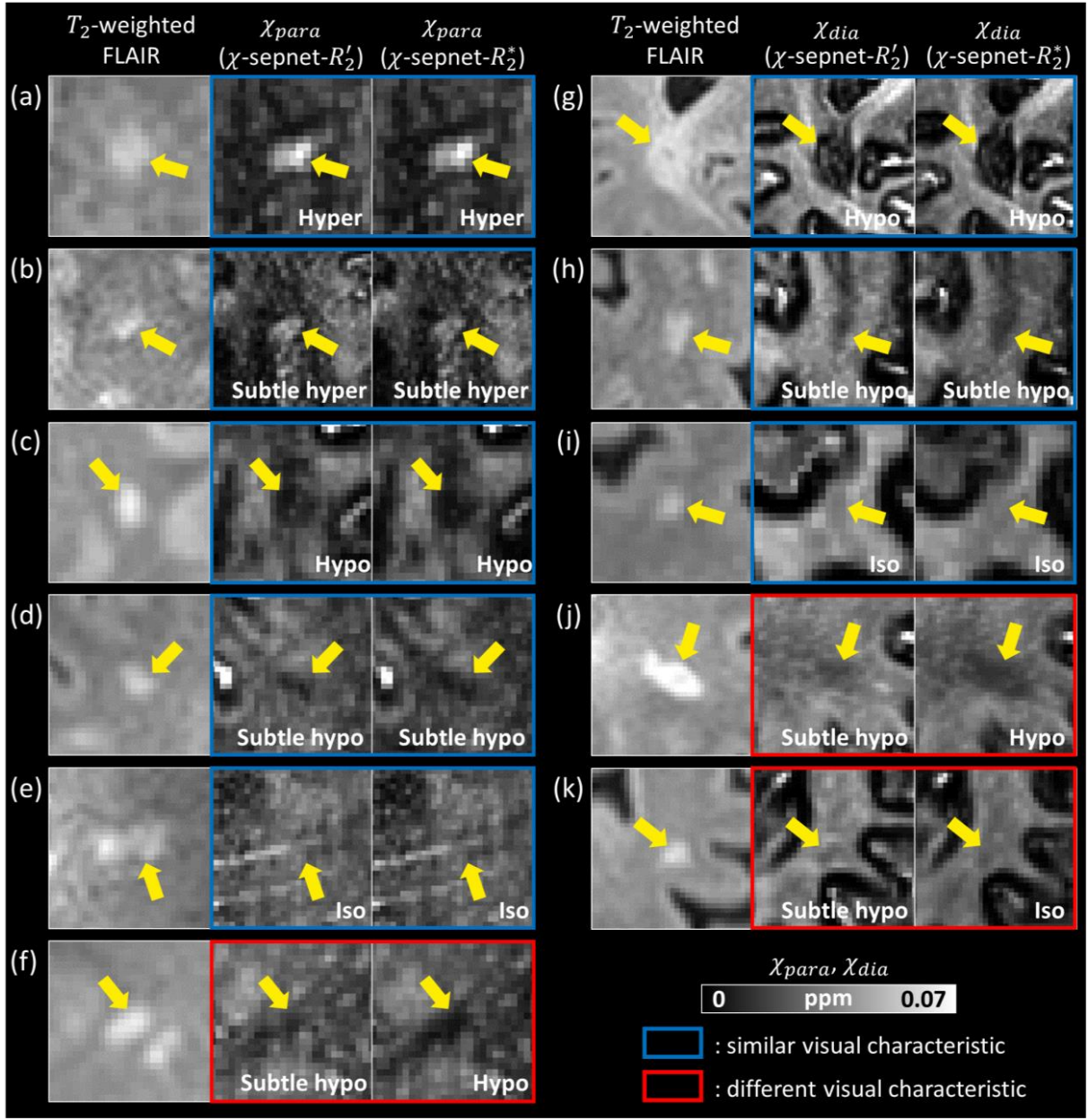

**Figure 6.** Sample MS lesion appearances in the $\chi_{para}$ and $\chi_{dia}$ maps generated by $\chi$-sepnet-$R_2'$ and $\chi$-sepnet-$R_2^*$. The lesion locations are marked by yellow arrows. The maps exhibiting similar visual characteristics are indicated by blue boxes, while those with different characteristics are indicated by red boxes. In $\chi_{para}$, most of the MS lesions exhibit similar contrasts in both maps, reporting (a) hyperintense, (b) subtle hyperintense, (c) hypointense, (d) subtle hypointense, and (e) isointense. Only one lesion out of 250 reveals slightly different visual characteristics (f). In the $\chi_{dia}$ maps, most MS lesions appear as (g) hypointense, (h) subtle hypointense, or (i) isointense. Only four lesions out of 250 were categorized as different lesion types, two of which are displayed in (j) and (k).

## 4. Discussion

In this study, we developed two χ-separation reconstruction pipelines, χ-sepnet-$R_2'$ and χ-sepnet-$R_2^*$, that successfully produced COSMOS-quality susceptibility source-separated maps in healthy subjects and MS patients. Both pipelines revealed high-quality images with good quantitative metrics. In particular, χ-sepnet-$R_2^*$ achieved good image quality and quantitative accuracy even without $R_2$ measurement, substantially improving the applicability of χ-separation in clinical and neuroscientific research.

In this work, multi-echo data processed QSMnet (QSMnet$_{ME}$) is developed to replace the original QSMnet, which utilized single-echo data. The metrics of QSMnet$_{ME}$ within the brain mask excluding CSF and vessels (pSNR: 33.3 ± 0.9 dB, NRMSE: 48.0 ± 4.3%, HFEN: 36.8 ± 2.8%, SSIM: 0.95 ± 0.01; Fig. 2b) and with only the brain mask (pSNR: 32.5 ± 0.9 dB, NRMSE: 47.1 ± 3.9%, HFEN: 36.1 ± 2.6%, SSIM: 0.95 ± 0.01) were slightly better than those of QSMnet (pSNR: 37.4 ± 1.1 dB, NRMSE: 50.0 ± 4.5%, HFEN: 48.5 ± 5.9%, SSIM: 0.91 ± 0.02) (Yoon et al., 2018) and QSMnet+ (pSNR: 37.3 ± 1.0 dB, NRMSE: 50.7 ± 4.8%, HFEN: 49.2 ± 6.1%, SSIM: 0.91 ± 0.02) (Jung et al., 2020) except for pSNR. The slight improvement may be from the larger number of head orientations used in the training dataset of QSMnet$_{ME}$.

R2PRIMEnet was proposed to infer $R_2'$ from $R_2^*$, which was motivated by the fact that $R_2^*$ contains information related to $R_2'$. Similar contrast conversion tasks using deep learning have been proposed previously (Hagiwara et al., 2019; Moya-Sáez et al., 2021; Tanenbaum et al., 2023). Furthermore, such a network was recently approved by a regulatory agency (Tanenbaum et al., 2023), supporting the reliability of such networks. Our quantitative metrics (pSNR: 42.7 ± 4.2 dB, NRMSE: 25.3 ± 1.9%, HFEN: 44.9 ± 5.2%, SSIM: 0.949 ± 0.006; Fig. 2d), which are better than those of QSMnet$_{ME}$ in pSNR and NRMSE, suggest that estimating $R_2'$ from $R_2^*$ is feasible.

When compared to the χ-sep-COSMOS results, maps from χ-sepnet based pipelines seem to provide less noisy results (Fig. 3-4). This may be because a network might focus on structural information rather than noise. Additionally, χ-sep-COSMOS may have encountered registration issues when processing data with multiple head orientations and multi-echo SE data that had thicker slices than the multi-echo GRE data.

When comparing the metrics of χ-sepnet-$R_2'$ ($\chi_{para}$ pSNR: 39.1 ± 1.0 dB, NRMSE: 38.1 ± 2.8%, HFEN: 47.2 ± 4.2%, SSIM: 0.940 ± 0.007; $\chi_{dia}$ pSNR: 39.5 ± 0.9 dB, NRMSE: 35.7 ± 2.2%, HFEN: 48.4 ± 2.7%, SSIM: 0.937 ± 0.008) and χ-sepnet-$R_2^*$ ($\chi_{para}$ pSNR: 38.4 ± 0.8 dB, NRMSE: 41.6 ± 2.4%, HFEN: 55.3 ± 5.1%, SSIM: 0.927 ± 0.007; $\chi_{dia}$ pSNR: 38.6 ± 0.8 dB, NRMSE: 39.4 ± 1.8%, HFEN: 57.7 ± 2.1%, SSIM: 0.922 ± 0.008) to those of QSMnet$_{ME}$ (pSNR: 33.3 ± 0.9 dB, NRMSE: 48.0 ± 4.3%, HFEN: 36.8 ± 2.8%, SSIM: 0.950 ± 0.011), they demonstrated improved performances in pSNR and NRMSE, although HFEN and SSIM of χ-sepnet-$R_2'$ and χ-sepnet-$R_2^*$ were somewhat inferior to those of QSMnet$_{ME}$. Overall, these suggested high-quality of the results of χ-sepnet-$R_2'$ and χ-sepnet-$R_2^*$.

In the MS lesion study, χ-sepnet-$R_2^*$ produced comparable lesion characteristic results to χ-sepnet-$R_2'$, confirming its utility (Tables 2, 3, and Figure 6). One limitation of lesion studies is that no gold-standard map exists for the patient data. Hence, one has to be cautious in interpreting the results. Additionally, the $R_2$ maps were obtained using DE-TSE instead of multi-echo SE, potentially introducing minor differences. Further investigation is required to study patients with other diseases that introduce different susceptibility characteristics.

Although the proposed pipelines demonstrate superior performance compared to the other conventional $\chi$-separation methods, they are subject to the intrinsic limitations of $\chi$-separation: large vessels introduce artifacts due to flow and non-local effects (Lee et al., 2024), and the models ignore susceptibility anisotropy (Lee et al., 2010), impeding the accurate reconstruction. Further efforts are necessary to improve the $\chi$-separation model.

The cascading of multiple networks, as utilized in our pipelines, can potentially propagate errors. For example, $\chi$-sepnet in the $\chi$-sepnet-$R_2^*$ pipeline has inputs that are outputs of QSMnet$_{ME}$ and R2PRIMEnet. Therefore, errors from QSMnet$_{ME}$ and R2PRIMEnet can be propagated into the reconstruction results of $\chi$-sepnet. While the design and evaluation of a single neural network that integrates these multiple networks could address this issue, it would require careful consideration of factors such as network capacity, interpretability, and the availability of sufficient training data. Additionally, QSMnet$_{ME}$ is limited by its tendency to learn a dipole kernel specific to the resolution of the training dataset, which restricts its performance to that particular resolution (Jung et al., 2022). Consequently, errors arising from resolutions other than the training data may be transmitted to $\chi$-sepnet. However, by using a resolution generalization method, this limitation of QSMnet$_{ME}$ can be partially overcome, and $\chi$-sepnet can effectively extend its resolution generalization ability as demonstrated in our results (Supplementary Section 4).

The proposed pipelines will encounter difficulties when processing out-of-distribution data from the trained datasets. For instance, in lesions with high susceptibility values, the pipelines may produce maps with underestimated values as demonstrated in QSMnet+ (Jung et al., 2020). Nevertheless, translating the scaling augmentation approach of QSMnet+ to R2PRIMEnet is not straightforward because there is no existing model that directly produces $R_2'$ from $R_2^*$. Hence, further research is necessary to develop a generalized $\chi$-sepnet.

This work has been partially presented at the 2021 Annual Meeting of the International Society of Magnetic Resonance in Medicine (Kim et al., 2022). It is important to note that $\chi$-sepnet mentioned in the abstract differs from the proposed pipelines in terms of the number of datasets used for the training and the inclusion of R2PRIMEnet.

## 5. Conclusion

In conclusion, we proposed two deep learning-powered susceptibility source separation pipelines, $\chi$-sepnet-$R_2'$ and $\chi$-sepnet-$R_2^*$. Leveraging the capabilities of the deep neural networks, $\chi$-sepnet-$R_2'$ successfully reconstructed COSMOS-quality paramagnetic and diamagnetic susceptibility source maps using single-head orientation multi-echo GRE data and multi-echo SE data. Furthermore, $\chi$-sepnet-$R_2^*$ produced $\chi$-separation maps with comparable quality using only multi-echo GRE data. This advancement significantly lowers the burden of data acquisition, thereby enhancing the clinical availability and impact of $\chi$-separation.

## Acknowledgments

This work was supported by the National Research Foundation of Korea (NRF-2022R1A4A1030579), NRF funded by the Korean government (MSIT) (RS-2024-00349509), Institute of Information & communications Technology Planning & Evaluation (IITP) under the artificial intelligence semiconductor support program to nurture the best talents (IITP-2023-RS-2023-00256081), INMC, IOER, and AI-Bio Research at Seoul National University.

## Conflict of interest statement

The authors have no conflicts of interest to disclose.

## Data availability statement

The proposed methods are available at https://github.com/SNU-LIST/chi_sepnet.
Chi-separation toolbox, including proposed deep learning-based reconstruction, is available at https://github.com/SNU-LIST/chi-separation.
The data used in this study are available on request from the corresponding author and can be shared following Institutional Review Board (IRB) approval due to privacy or ethical restrictions.

## References

Bender, B., Klose, U., 2010. The in vivo influence of white matter fiber orientation towards B0 on T2* in the human brain. NMR in Biomedicine 23, 1071-1076.

Biondetti, E., Karsa, A., Thomas, D.L., Shmueli, K., 2020. Investigating the accuracy and precision of TE-dependent versus multi-echo QSM using Laplacian-based methods at 3 T. Magnetic resonance in medicine 84, 3040-3053.

Bollmann, S., Rasmussen, K.G.B., Kristensen, M., Blendal, R.G., Østergaard, L.R., Plocharski, M., O'Brien, K., Langkammer, C., Janke, A., Barth, M., 2019. DeepQSM-using deep learning to solve the dipole inversion for quantitative susceptibility mapping. Neuroimage 195, 373-383.

Chen, J., Gong, N.-J., Chaim, K.T., Otaduy, M.C.G., Liu, C., 2021. Decompose quantitative susceptibility mapping (QSM) to sub-voxel diamagnetic and paramagnetic components based on gradient-echo MRI data. Neuroimage 242.

Committee, Q.S.M.C.O., Bilgic, B., Costagli, M., Chan, K.S., Duyn, J., Langkammer, C., Lee, J., Li, X., Liu, C., Marques, J.P., Milovic, C., Robinson, S.D., Schweser, F., Shmueli, K., Spincemaille, P., Straub, S., van Zijl, P., Wang, Y., Group, I.E.-M.T.P.S., 2024. Recommended implementation of quantitative susceptibility mapping for clinical research in the brain: A consensus of the ISMRM electro-magnetic tissue properties study group. Magn Reson Med 91, 1834-1862.

Connor, J., Menzies, S., St. Martin, S., Mufson, E., 1992. A histochemical study of iron, transferrin, and ferritin in Alzheimer's diseased brains. Journal of neuroscience research 31, 75-83.

de Rochefort, L., Brown, R., Prince, M.R., Wang, Y., 2008. Quantitative MR susceptibility mapping using piece-wise constant regularized inversion of the magnetic field. Magn Reson Med 60, 1003-1009.

Dexter, D., Wells, F., Agid, F., Agid, Y., Lees, A., Jenner, P., Marsden, C., 1987. Increased nigral iron content in postmortem parkinsonian brain. The Lancet 330, 1219-1220.

Dimov, A.V., Nguyen, T.D., Gillen, K.M., Marcille, M., Spincemaille, P., Pitt, D., Gauthier, S.A., Wang, Y., 2022. Susceptibility source separation from gradient echo data using magnitude decay modeling. J Neuroimaging 32, 852-859.

Dymerska, B., Eckstein, K., Bachrata, B., Siow, B., Trattnig, S., Shmueli, K., Robinson, S.D., 2021. Phase unwrapping with a rapid opensource minimum spanning tree algorithm (ROMEO). Magn Reson Med 85, 2294-2308.

Egger, C., Opfer, R., Wang, C., Kepp, T., Sormani, M.P., Spies, L., Barnett, M., Schippling, S., 2017. MRI FLAIR lesion segmentation in multiple sclerosis: Does automated segmentation hold up with manual annotation? Neuroimage Clin 13, 264-270.

Emmerich, J., Bachert, P., Ladd, M.E., Straub, S., 2021. On the separation of susceptibility sources in quantitative susceptibility mapping: Theory and phantom validation with an in vivo application to multiple sclerosis lesions of different age. Journal of Magnetic Resonance 330, 107033.

Fukunaga, M., Li, T.-Q., van Gelderen, P., de Zwart, J.A., Shmueli, K., Yao, B., Lee, J., Maric, D., Aronova, M.A., Zhang, G., 2010. Layer-specific variation of iron content in cerebral cortex as a source of MRI contrast. Proceedings of the National Academy of Sciences 107, 3834-3839.

Hagiwara, A., Otsuka, Y., Hori, M., Tachibana, Y., Yokoyama, K., Fujita, S., Andica, C., Kamagata, K., Irie, R., Koshino, S., 2019. Improving the quality of synthetic FLAIR images with deep learning using a conditional generative adversarial network for pixel-by-pixel image translation. American Journal of Neuroradiology 40, 224-230.

Jang, M., Dimov, A.V., Kapse, K., Murnick, J., Grinspan, Z., Wu, A., Choudhury, A.R., Wang, Y., Spincemaille, P., Nguyen, T.D., 2024. Quantitative susceptibility mapping with source separation in normal brain development of newborns. American Journal of Neuroradiology.

Ji, S., Jang, J., Kim, M., Lee, H., Kim, W., Lee, J., Shin, H.G., 2024. Comparison between R2'-based and R2*-based chi-separation methods: A clinical evaluation in individuals with multiple sclerosis. NMR Biomed, e5167.

Ji, S., Park, J., Shin, H.-G., Youn, J., Kim, M., Lee, J., 2023. Successfulgeneralization for data with higher or lower resolution than training data resolution in deep learning powered QSM reconstruction. Joint Annual Meeting ISMRM-ESMRMB & ISMRT 32nd Annual Meeting, Toronto, Canada.

Jung, W., Bollmann, S., Lee, J., 2022. Overview of quantitative susceptibility mapping using deep learning: Current status, challenges and opportunities. NMR Biomed 35, e4292.

Jung, W., Yoon, J., Ji, S., Choi, J.Y., Kim, J.M., Nam, Y., Kim, E.Y., Lee, J., 2020. Exploring linearity of deep neural network trained QSM: QSMnet(). Neuroimage 211, 116619.

Kan, H., Uchida, Y., Kawaguchi, S., Kasai, H., Hiwatashi, A., Ueki, Y., 2024. Quantitative Susceptibility Mapping for Susceptibility Source Separation with Adaptive Relaxometric Constant Estimation (QSM-ARCS) from Solely Gradient-Echo Data. Neuroimage, 120676.

Kim, M., Shin, H.-G., Oh, C., Jeong, H., Ji, S., An, H., Kim, J., Jang, J., Bilgic, B., Lee, J., 2022. Chi-sepnet: Susceptibility source separation using deep neural network. Joint Annual Meeting ISMRM-ESMRMB & ISMRT 31st Annual Meeting, London, UK.

Kim, T., Ji, S., Min, K., Youn, J., Kim, M., Kim, J., Lee, J., 2024. Physics-informed vessel segmentation for χ-separation (chi-separation). Joint Annual Meeting ISMRM-ESMRMB & ISMRT 33rd Annual Meeting, Singapore.

Kim, W., Shin, H.G., Lee, H., Park, D., Kang, J., Nam, Y., Lee, J., Jang, J., 2023. chi-Separation Imaging for Diagnosis of Multiple Sclerosis versus Neuromyelitis Optica Spectrum Disorder. Radiology 307, e220941.

Lee, J., Ji, S., Oh, S.-H., 2024. So You Want to Image Myelin Using MRI: Magnetic Susceptibility Source Separation for Myelin Imaging. Magnetic Resonance in Medical Sciences, rev. 2024-0001.

Lee, J., Nam, Y., Choi, J.Y., Shin, H.G., Hwang, T., Lee, J., 2017. Separating positive and negative susceptibility sources in QSM. Proceedings of the ISMRM 25th Annual Meeting & Exhibition, pp. 22-27.

Lee, J., Shmueli, K., Fukunaga, M., Van Gelderen, P., Merkle, H., Silva, A.C., Duyn, J.H., 2010. Sensitivity of MRI resonance frequency to the orientation of brain tissue microstructure. Proceedings of the National Academy of Sciences 107, 5130-5135.

Lee, J., van Gelderen, P., Kuo, L.-W., Merkle, H., Silva, A.C., Duyn, J.H., 2011. T2*-based fiber orientation mapping. Neuroimage 57, 225-234.

Lee, S., Shin, H.G., Kim, M., Lee, J., 2023. Depth-wise profiles of iron and myelin in the cortex and white matter using chi-separation: A preliminary study. Neuroimage 273, 120058.

Li, W., Wang, N., Yu, F., Han, H., Cao, W., Romero, R., Tantiwongkosi, B., Duong, T.Q., Liu, C., 2015. A method for estimating and removing streaking artifacts in quantitative susceptibility mapping. Neuroimage 108, 111-122.

Li, Z., Feng, R., Liu, Q., Feng, J., Lao, G., Zhang, M., Li, J., Zhang, Y., Wei, H., 2023. APART-QSM: An improved sub-voxel quantitative susceptibility mapping for susceptibility source separation using an iterative data fitting method. Neuroimage 274, 120148.

Liu, T., Liu, J., de Rochefort, L., Spincemaille, P., Khalidov, I., Ledoux, J.R., Wang, Y., 2011. Morphology enabled dipole inversion (MEDI) from a single-angle acquisition: comparison with COSMOS in human brain imaging. Magn Reson Med 66, 777-783.

Liu, T., Spincemaille, P., de Rochefort, L., Kressler, B., Wang, Y., 2009. Calculation of susceptibility through multiple orientation sampling (COSMOS): a method for conditioning the inverse problem from measured magnetic field map to susceptibility source image in MRI. Magn Reson Med 61, 196-204.

Liu, Z., Spincemaille, P., Yao, Y., Zhang, Y., Wang, Y., 2018. MEDI+0: Morphology enabled dipole inversion with automatic uniform cerebrospinal fluid zero reference for quantitative susceptibility mapping. Magn Reson Med 79, 2795-2803.

Martino, G., Hartung, H.-P., 1999. Immunopathogenesis of multiple sclerosis: the role of T cells. Current opinion in neurology 12, 309-321.

McPhee, K.C., Wilman, A.H., 2015. T2 quantification from only proton density and T2-weighted MRI by modelling actual refocusing angles. Neuroimage 118, 642-650.

Min, K., Sohn, B., Kim, W.J., Park, C.J., Song, S., Shin, D.H., Chang, K.W., Shin, N.-Y., Kim, M., Shin, H.-G., 2023. A human brain atlas of chi-separation for normative iron and myelin distributions. arXiv preprint arXiv:2311.04468.

Moya-Sáez, E., Peña-Nogales, Ó., de Luis-García, R., Alberola-López, C., 2021. A deep learning approach for synthetic MRI based on two routine sequences and training with synthetic data. Computer Methods and Programs in Biomedicine 210, 106371.

Müller, J., Lu, P.-J., Cagol, A., Ruberte, E., Shin, H.-G., Ocampo-Pineda, M., Chen, X., Tsagkas, C., Barakovic, M., Galbusera, R., 2024. Quantifying Remyelination Using χ-Separation in White Matter and Cortical Multiple Sclerosis Lesions. Neurology 103, e209604.

Naidich, T.P., Castillo, M., Cha, S., Smirniotopoulos, J.G., 2012. Imaging of the brain E-book: Expert radiology series. Elsevier Health Sciences.

Nakashima, M., Kan, H., Kawai, T., Matsumoto, K., Kawaguchi, T., Uchida, Y., Matsukawa, N., Hiwatashi, A., 2024. Quantitative susceptibility mapping analyses of white matter in Parkinson's disease using susceptibility separation technique. Parkinsonism & Related Disorders, 107135.

Oh, S.H., Kim, Y.B., Cho, Z.H., Lee, J., 2013. Origin of B0 orientation dependent R2(*) (=1/T2(*)) in white matter. Neuroimage 73, 71-79.

Paszke, A., Gross, S., Massa, F., Lerer, A., Bradbury, J., Chanan, G., Killeen, T., Lin, Z., Gimelshein, N., Antiga, L., 2019. Pytorch: An imperative style, high-performance deep learning library. Advances in Neural Information Processing Systems 32.

Pei, M., Nguyen, T.D., Thimmappa, N.D., Salustri, C., Dong, F., Cooper, M.A., Li, J., Prince, M.R., Wang, Y., 2015. Algorithm for fast monoexponential fitting based on Auto-Regression on Linear Operations (ARLO) of data. Magn Reson Med 73, 843-850.

Robinson, S., Grabner, G., Witoszynskyj, S., Trattnig, S., 2011. Combining phase images from multi-channel RF coils using 3D phase offset maps derived from a dual-echo scan. Magnetic resonance in medicine 65, 1638-1648.

Schaltenbrand, G., 1977. Atlas for stereotaxy of the human brain. Georg Thieme Publishers.

Schweser, F., Deistung, A., Lehr, B.W., Reichenbach, J.R., 2011. Quantitative imaging of intrinsic magnetic tissue properties using MRI signal phase: an approach to in vivo brain iron metabolism? Neuroimage 54, 2789-2807.

Shin, H.G., Lee, J., Yun, Y.H., Yoo, S.H., Jang, J., Oh, S.H., Nam, Y., Jung, S., Kim, S., Fukunaga, M., Kim, W., Choi, H.J., Lee, J., 2021. chi-separation: Magnetic susceptibility source separation toward iron and myelin mapping in the brain. Neuroimage 240, 118371.

Shmueli, K., de Zwart, J.A., van Gelderen, P., Li, T.Q., Dodd, S.J., Duyn, J.H., 2009. Magnetic susceptibility mapping of brain tissue in vivo using MRI phase data. Magn Reson Med 62, 1510-1522.

Smith, S.M., Jenkinson, M., Woolrich, M.W., Beckmann, C.F., Behrens, T.E., Johansen-Berg, H., Bannister, P.R., De Luca, M., Drobnjak, I., Flitney, D.E., Niazy, R.K., Saunders, J., Vickers, J., Zhang, Y., De Stefano, N., Brady, J.M., Matthews, P.M., 2004. Advances in functional and structural MR image analysis and implementation as FSL. Neuroimage 23 Suppl 1, S208-219.

Stuber, C., Morawski, M., Schafer, A., Labadie, C., Wahnert, M., Leuze, C., Streicher, M., Barapatre, N., Reimann, K., Geyer, S., Spemann, D., Turner, R., 2014. Myelin and iron concentration in the human brain: a quantitative study of MRI contrast. Neuroimage 93 Pt 1, 95-106.

Tanenbaum, L., Bash, S., Zaharchuk, G., Shankaranarayanan, A., Chamberlain, R., Wintermark, M., Beaulieu, C., Novick, M., Wang, L., 2023. Deep learning–generated synthetic MR imaging STIR spine images are superior in image quality and diagnostically equivalent to conventional STIR: a multicenter, multireader trial. American Journal of Neuroradiology 44, 987-993.

Wang, Y., Liu, T., 2015. Quantitative susceptibility mapping (QSM): Decoding MRI data for a tissue magnetic biomarker. Magn Reson Med 73, 82-101.

Wingerchuk, D.M., Lennon, V.A., Lucchinetti, C.F., Pittock, S.J., Weinshenker, B.G., 2007. The spectrum of neuromyelitis optica. The Lancet Neurology 6, 805-815.

Wu, B., Li, W., Avram, A.V., Gho, S.-M., Liu, C., 2012a. Fast and tissue-optimized mapping of magnetic susceptibility and T2* with multi-echo and multi-shot spirals. Neuroimage 59, 297-305.

Wu, B., Li, W., Guidon, A., Liu, C., 2012b. Whole brain susceptibility mapping using compressed sensing. Magn Reson Med 67, 137-147.

Yoon, J., Gong, E., Chatnuntawech, I., Bilgic, B., Lee, J., Jung, W., Ko, J., Jung, H., Setsompop, K., Zaharchuk, G., Kim, E.Y., Pauly, J., Lee, J., 2018. Quantitative susceptibility mapping using deep neural network: QSMnet. Neuroimage 179, 199-206.

Zhu, Z., Naji, N., Esfahani, J.H., Snyder, J., Seres, P., Emery, D.J., Noga, M., Blevins, G., Smyth, P., Wilman, A.H., 2024. MR Susceptibility Separation for Quantifying Lesion Paramagnetic and Diamagnetic Evolution in Relapsing-Remitting Multiple Sclerosis. J Magn Reson Imaging.

# Supplementary information

## 1. 3D U-net architectures for χ-sepnet, QSMnet$_{ME}$, and R2PRIMEnet

The network architecture for χ-sepnet, QSMnet$_{ME}$, and R2PRIMEnet was based on the 3D U-net used in QSMnet (Yoon et al., 2018). QSMnet$_{ME}$ and R2PRIMEnet followed the same design as QSMnet, while χ-sepnet had input and output channel numbers modified to 3 and 2, respectively (Supplementary Fig. 1). The details of the 3D U-net architecture are illustrated in Supplementary Figure 1. The 3D U-net consisted of 19 convolutional layers (18 with a kernel size = $3 \times 3 \times 3$ and one with a kernel size of $1 \times 1 \times 1$), 18 batch normalization layers, 18 ReLU layers, 4 max-pooling layers (kernel size = $2 \times 2 \times 2$), 4 deconvolutional layers (kernel size = $2 \times 2 \times 2$), and 4 skip connections to preserve spatial information.

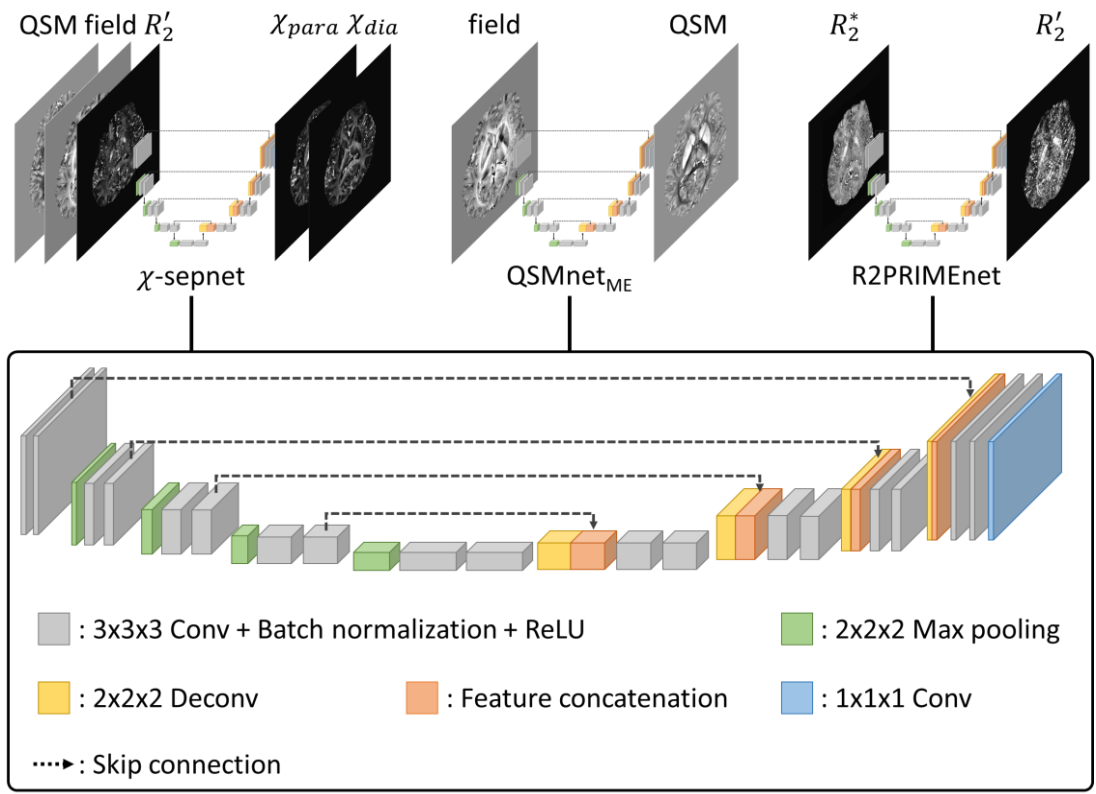

**Supplementary Figure 1**. Network architectures of χ-sepnet, QSMnet$_{ME}$, and R2PRIMEnet.

*2. ROI analysis in healthy subjects*

The ROI analysis was performed with the susceptibility source-separated maps from χ-sep-COSMOS, χ-sepnet-$R_2'$, χ-sepnet-$R_2^*$, χ-sep-iLSQR, and χ-sep-MEDI in the healthy subject test dataset. We employed ROI masks from the χ-separation atlas (Min et al., 2023), which included masks of gray matters (caudate nucleus [CN], putamen [Put], globus pallidus [GP], nucleus accumbens [NA], substantia nigra [SN], red nucleus [RN], ventral pallidum [VP], subthalamic nucleus [STN], medial thalamic nuclei [MTh], lateral thalamic nuclei [LTh], pulvinar [Pul]), and white matters (genu of corpus callosum [GN], body of corpus callosum [BD], splenium of corpus callosum [SP], cerebral peduncle [CP], anterior limb of internal capsule [ALIC], posterior limb of internal capsule [PLIC], retrolenticular part of internal capsule [RPIC], anterior corona radiata [ACR], superior corona radiata [SCR], posterior corona radiata [PCR], posterior thalamic radiation [PTR], sagittal stratum [SS], superior longitudinal fasciculus [SLF]). To transform the χ-separation ROI masks from the template, which was in the MNI space, to the space of an individual subject, we followed the process described in the χ-separation atlas paper (Min et al., 2023): First, the T1-weighted image of the healthy subject was registered to the COSMOS QSM map (Liu et al., 2009) of the same subject using ANTs (https://github.com/ANTsX/ANTs). The registered T1-weighted image and COSMOS QSM map were linearly combined to create a hybrid image. Then, the hybrid image was transformed to the MNI space, and the inverse affine transformation matrix was calculated. Finally, using this inverse affine transform matrix, the χ-separation atlas ROI masks were transformed to the space of the subject. The transformed ROI masks were used to report the mean and standard deviation of the susceptibility values of the ROIs (Supplementary Fig. 2).

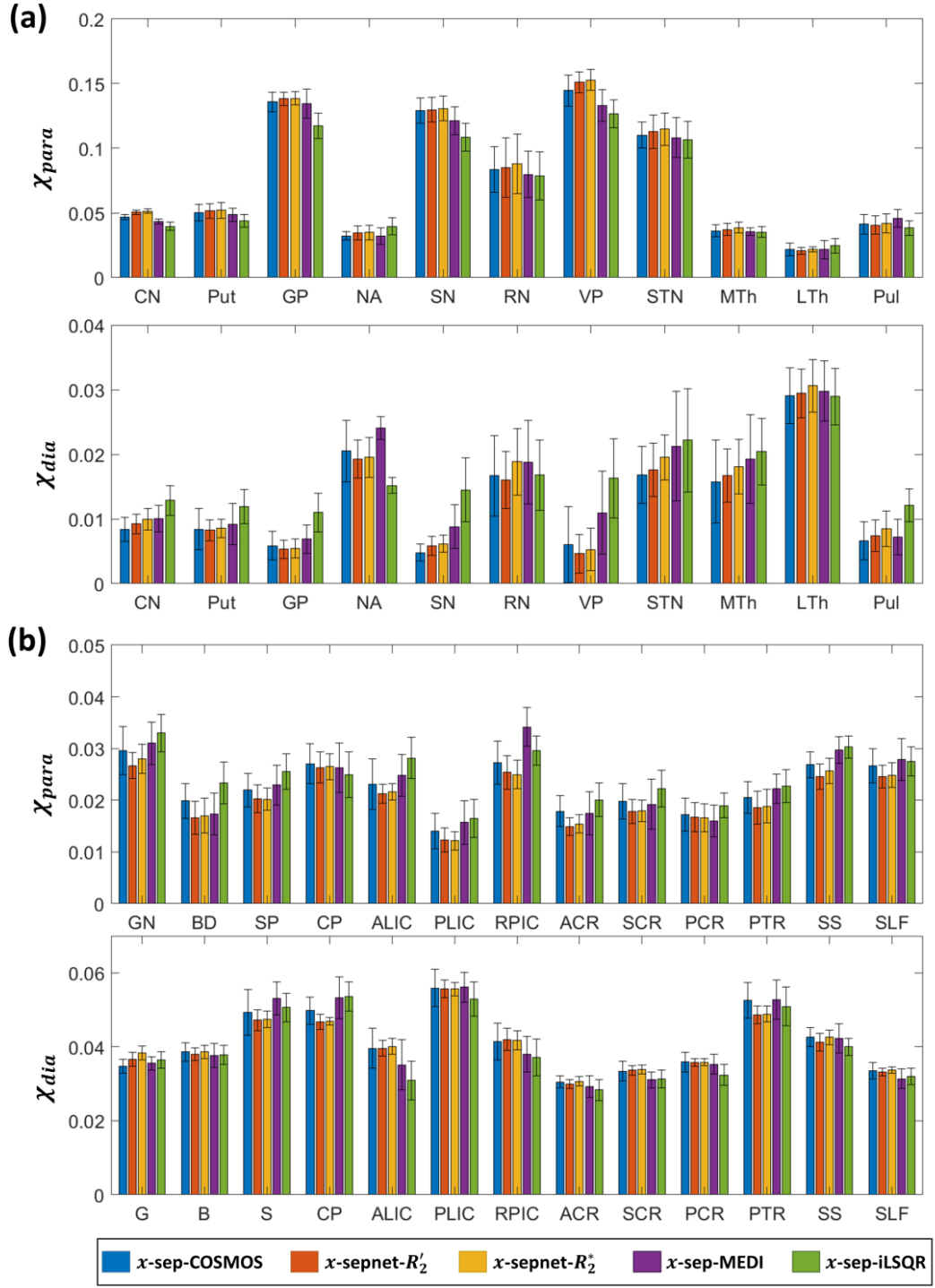

**Supplementary Figure 2**. ROI analysis results reporting the mean and standard deviation over the six test healthy subjects. A total of 24 ROI masks are used for the analysis, including (a) gray matter and (b) white matter regions.

*3. Results of input ablation study*

   The results of the input ablation study that compare the performance of $\chi$-sepnet with the models that excluded either the QSM input (i.e., $\chi$-sepnet trained with field and $R_2'$ only) or the local field input (i.e., $\chi$-sepnet trained with QSM and $R_2'$ only), are shown in Supplementary Fig. 3 and Supplementary Table 1. The proposed model shows the overall best performance although all three models show almost the same performance.

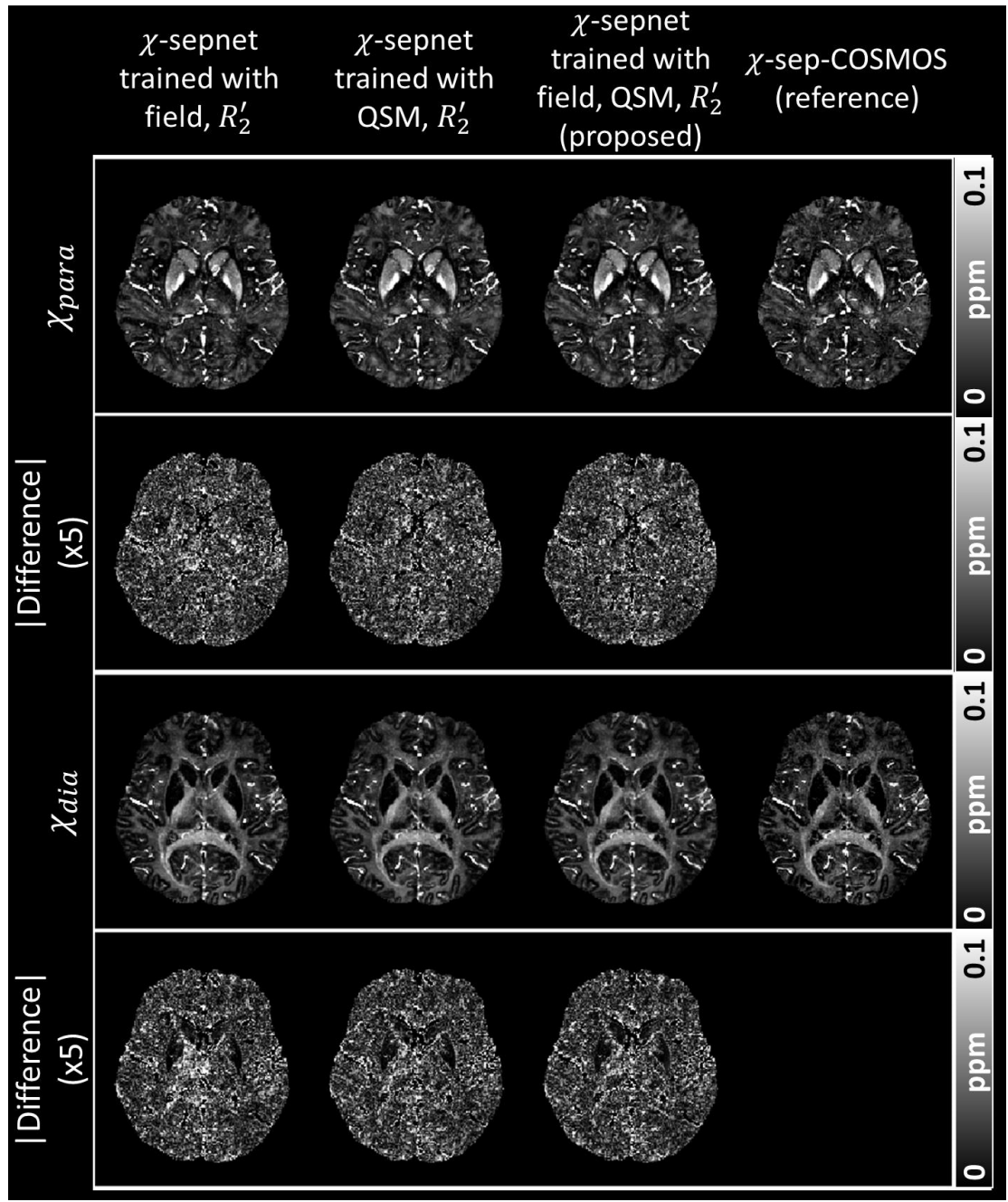

**Supplementary Figure 3**. Results of the input ablation study of $\chi$-sepnet.

**Supplementary Table 1.** Quantitative metric results of the $\chi$-sepnet-$R_2'$ pipeline input ablation study.

|  |  | $\chi$-sepnet trained with field, $R_2'$ | $\chi$-sepnet trained with QSM, $R_2'$ | $\chi$-sepnet trained with field, QSM, $R_2'$ (proposed) |
|---|---|---|---|---|
| $\chi_{\text{para}}$ | pSNR($\uparrow$) | 39.0 $\pm$ 1.0 dB | 39.1 $\pm$ 1.0 dB | **39.1 $\pm$ 1.0 dB** |
|  | NRMSE($\downarrow$) | 38.6 $\pm$ 2.8% | 38.2 $\pm$ 2.8% | **38.1 $\pm$ 2.8%** |
|  | HFEN($\downarrow$) | 50.9 $\pm$ 3.5% | **47.2 $\pm$ 4.1%** | 47.2 $\pm$ 4.2% |
|  | SSIM($\uparrow$) | 0.938 $\pm$ 0.008 | **0.940 $\pm$ 0.008** | 0.940 $\pm$ 0.007 |
| $\chi_{\text{dia}}$ | pSNR($\uparrow$) | 39.3 $\pm$ 0.9 dB | 39.4 $\pm$ 0.9 dB | **39.5 $\pm$ 0.9 dB** |
|  | NRMSE($\downarrow$) | 36.3 $\pm$ 2.2% | 35.8 $\pm$ 2.1% | **35.7 $\pm$ 2.2%** |
|  | HFEN($\downarrow$) | 51.9 $\pm$ 3.1% | 48.5 $\pm$ 2.7% | **48.4 $\pm$ 2.7%** |
|  | SSIM($\uparrow$) | 0.936 $\pm$ 0.008 | 0.937 $\pm$ 0.008 | **0.937 $\pm$ 0.008** |

*4. Results of resolution generalization study*

To assess the resolution generalization capability of $\chi$-sepnet, three distinct resolutions compared to the trained dataset (1 mm isotropic resolution) were tested: 2 × 2 × 2 mm³, 1 × 1 × 3 mm³, and 0.7 × 0.7 × 0.7 mm³. The proposed model shows the overall best performance although all three models show comparable performance in isotropic resolution cases (2 × 2 × 2 mm³ and 0.7 × 0.7 × 0.7 mm³) (Supplementary Fig. 4 and 6). The model that employs the local field and $R_2'$ maps reveals the poorest performance on anisotropic resolution case (1 × 1 × 3 mm³) compared to the other models (Supplementary Fig. 5). This result, consistent with the findings of the previous study (Jung et al., 2022), indicates that a network trained solely with the field map and $R_2'$ map was limited to learning the dipole kernel specific to the resolution of the training dataset, thereby exhibiting poor resolution generalization capability.

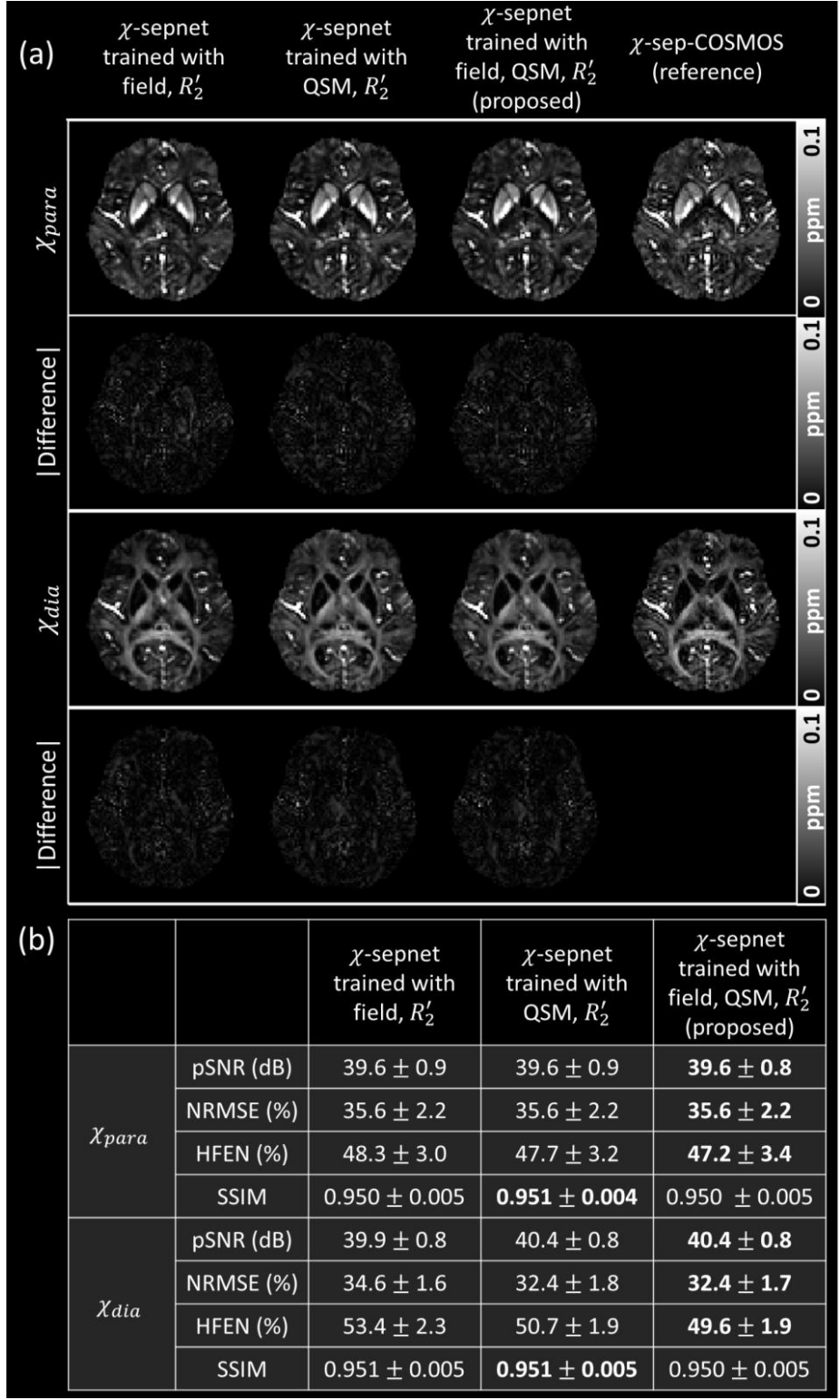

**Supplementary Figure 4**. Results of the resolution generalization study of χ-sepnet in 2 × 2 × 2 mm$^3$ resolution. (a) $\chi_{para}$ and $\chi_{dia}$ maps from χ-sepnets and χ-sep-COSMOS are displayed (first row for $\chi_{para}$ and third row for $\chi_{dia}$) with the difference images (second and fourth rows). (b) Quantitative metrics.

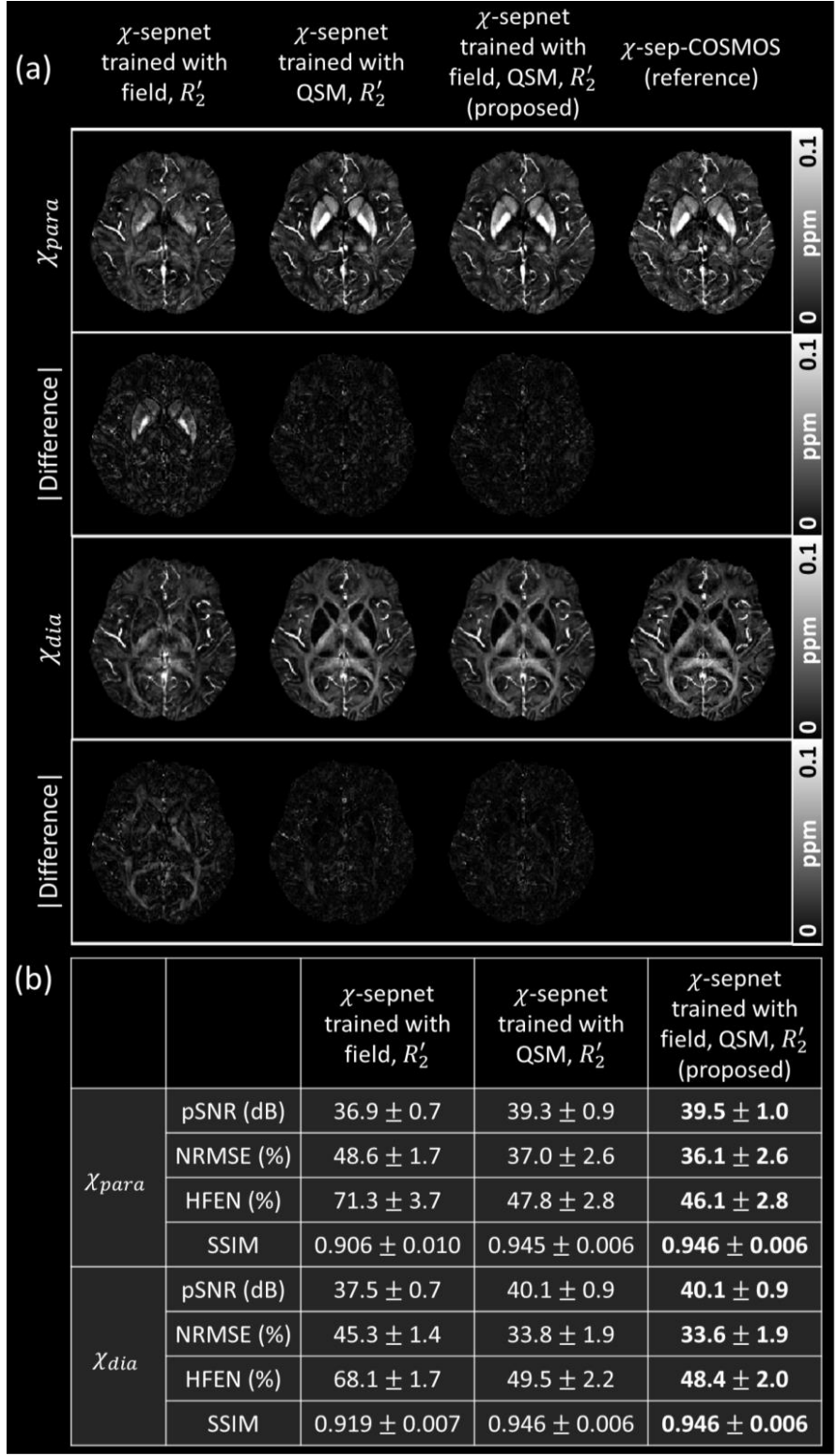

**Supplementary Figure 5.** Results of the resolution generalization study of χ-sepnet in $1 \times 1 \times 3$ mm$^3$ resolution. (a) $\chi_{para}$ and $\chi_{dia}$ maps from three χ-sepnets and χ-sep-COSMOS (first row for $\chi_{para}$ and third row for $\chi_{dia}$) with the difference images (second and fourth rows). (b) Quantitative metrics.

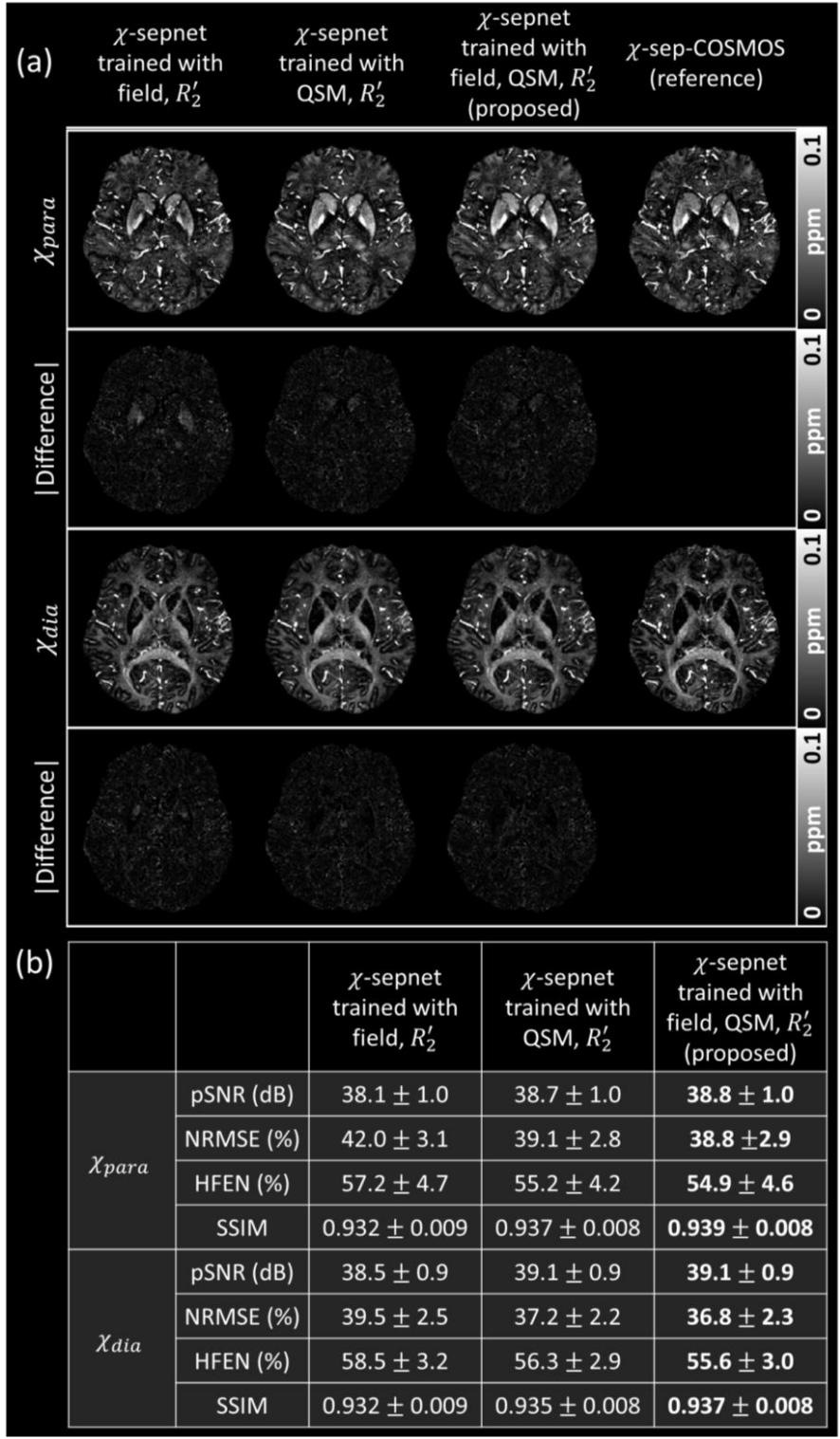

**Supplementary Figure 6**. Results of the resolution generalization study of χ-sepnet in 0.7 × 0.7 × 0.7 mm³ resolution. (a) $\chi_{para}$ and $\chi_{dia}$ maps from three χ-sepnets and χ-sep-COSMOS are displayed (first row for $\chi_{para}$ and third row for $\chi_{dia}$) with the difference images (second and fourth rows). (b) Quantitative metrics.

# References

Jung, W., Bollmann, S., Lee, J., 2022. Overview of quantitative susceptibility mapping using deep learning: Current status, challenges and opportunities. NMR Biomed 35, e4292.

Liu, T., Spincemaille, P., de Rochefort, L., Kressler, B., Wang, Y., 2009. Calculation of susceptibility through multiple orientation sampling (COSMOS): a method for conditioning the inverse problem from measured magnetic field map to susceptibility source image in MRI. Magn Reson Med 61, 196-204.

Min, K., Sohn, B., Kim, W.J., Park, C.J., Song, S., Shin, D.H., Chang, K.W., Shin, N.-Y., Kim, M., Shin, H.-G., 2023. A human brain atlas of chi-separation for normative iron and myelin distributions. arXiv preprint arXiv:2311.04468.

Yoon, J., Gong, E., Chatnuntawech, I., Bilgic, B., Lee, J., Jung, W., Ko, J., Jung, H., Setsompop, K., Zaharchuk, G., Kim, E.Y., Pauly, J., Lee, J., 2018. Quantitative susceptibility mapping using deep neural network: QSMnet. Neuroimage 179, 199-206.